\documentclass[10pt,journal,compsoc]{IEEEtran}

\usepackage{cite}
\usepackage{amsmath,amssymb,amsfonts}
\usepackage{gnuplottex}
\usepackage{algorithmic}
\usepackage{graphicx}
\graphicspath{{}}
\usepackage{caption}
\usepackage{subcaption}

\usepackage{textcomp}
\usepackage{microtype}
\usepackage[detect-all]{siunitx}
\DeclareSIUnit{\million}{\text{million}}
\DeclareSIUnit{\bits}{bits}
\DeclareSIUnit{\bit}{bit}
\DeclareSIUnit{\siTimes}{\ensuremath{\times}}
\newcommand{\SIadj}[2]{\num{#1}-\si{#2}}
\usepackage{xfrac}
\usepackage{xcolor}
\usepackage[normalem]{ulem}

\usepackage{csquotes}
\usepackage{enumitem}

\usepackage{booktabs}

\usepackage{tikz}
\usetikzlibrary{positioning}
\usetikzlibrary{backgrounds,scopes}
\usetikzlibrary{fit}
\usetikzlibrary{shapes, shapes.geometric}
\usetikzlibrary{matrix}
\usetikzlibrary{math}
\usetikzlibrary{calc}
\usetikzlibrary{arrows.meta}
\usetikzlibrary{decorations.pathreplacing}
\usetikzlibrary{external}
\usepackage{tikzpagenodes}

\tikzset{
    font={\fontsize{8pt}{9}\selectfont},
    arrow/.style={-latex}
    }

\usepackage{xparse}
\ExplSyntaxOn
\DeclareExpandableDocumentCommand{\convertlen}{ O{cm} m }
{
	\dim_to_decimal_in_unit:nn { #2 } { 1 #1 } cm
}
\ExplSyntaxOff

\usepackage{xspace}
\xspaceaddexceptions{()}
\usepackage[acronyms]{glossaries}

\newglossaryentry{ngram}{
	name={\emph{n}-gram},
	description={Text encoding building block for \emph{n} characters.}
	}

\newglossaryentry{vdd}{
	name={\ensuremath{V_{\text{DD}}}},
	description={supply voltage},
	symbol=Vdd,
	sort=vdd}
\newglossaryentry{vth}{
	name={\ensuremath{V_{\text{th}}}},
	description={threshold voltage},
	symbol=Vth,
	sort=vth}
\newglossaryentry{gnd}{
	name={GND},
	description={El. ground},
	symbol=GND}

\newglossaryentry{ntype}{
	name={n-type},
	description={pnp doted transistor. Other known names are nmos or nfet}}
\newglossaryentry{ptype}{
	name={p-type},
	description={npn doted transistor. Other known names are pmos or pfet}}
\newglossaryentry{ion}{
	name={\ensuremath{I_{\text{ON}}}},
	description={Current through transistors when it fully conducts},
	symbol=Ion,
	sort=ion}
\newglossaryentry{ioff}{
	name={\ensuremath{I_{\text{OFF}}}},
	description={Current through transistor when it is fully closed},
	symbol=Ioff
	sort=ioff}
\newglossaryentry{subSlope}{
	name={subthreshold-slope},
	description={Slope of \gls{transChar} in log scale below \gls{vth}}}
\newglossaryentry{transChar}{name={transfer characteristic},description={Name of the \iv curve}}

\newglossaryentry{vds}{
	name={\ensuremath{V_{\text{\textit{DS}}}}},
	description={Voltage difference between drain and source contact of a transistor},
	sort=vds}

\newglossaryentry{bitline}{name={bit line},description={Wires that connect columns of \acrshort{sram} cells}}

\newglossaryentry{q}{
	name={Q},
	description={Left inner node of a \acrshort{sram} cell},symbol=Q}
\newglossaryentry{qb}{
	name={QB},
	description={Right inner node of a \acrshort{sram} cell},symbol=QB}

\newacronym{dnn}{DNN}{Deep Neural Network}
\newacronym{tpu}{TPU}{Tensor Processing Unit}
\newacronym{hdc}{HDC}{Hyperdimensional Computing}
\newacronym{hd}{HD}{Hyperdimensional}
\newacronym{am}{AM}{Associative Memory}
\newacronym{im}{IM}{Item Memory}
\newacronym{nvm}{NVM}{non-volatile memory}

\newacronym{ml}{ML}{Match Line}
\newacronym{sl}{SL}{Select Line}
\newacronym{wl}{WL}{Word Line}
\newacronym{bl}{BL}{Bit Line}
\newacronym{blb}{BLB}{Bit Line Bar}
\newacronym{slb}{SLB}{Select Line Bar}

\newacronym{fet}{FET}{Field Effect Transistor}
\newacronym{mosfet}{MOSFET}{Metal Oxide Semiconductor \acrlong{fet}}
\newacronym{fefet}{FeFET}{Ferroelectric \gls{fet}}
\newacronym{fefinfet}{Fe-FinFET}{Ferroelectric \acrshort{finfet}}
\newacronym{ic}{IC}{Integrated Circuit}
\newacronym{ram}{RAM}{Random Access Memory}
\newacronym{cam}{CAM}{Content Addressable Memory}
\newacronym{tcam}{TCAM}{Ternary Content Addressable Memory}
\newacronym{sram}{SRAM}{Static Random Access Memory}
\newacronym{cmos}{CMOS}{Complementary MOS}
\newacronym{adc}{ADC}{Analog Digital Converter}
\newacronym{tdc}{TDC}{time-to-digital converter}
\newacronym{mlc}{MLC}{multi-level cell}

\newacronym{fpga}{FPGA}{Field Programmable Gate Array}
\newacronym{csrsa}{CSRSA}{clocked self-referenced sense amplifier}

\newacronym{fe}{FE}{ferroelectric}

\newacronym{stt}{STT-MRAM}{Spin-Transfer Torque Magnetic RAM}
\newacronym{rram}{ReRAM}{Resistive RAM}

\newacronym{finfet}{FinFET}{Fin \acrlong{fet}}
\newacronym{nc}{NC}{Negative Capacitance}
\newacronym{ncfet}{NCFET}{\acrlong{nc} \acrlong{fet}}

\newacronym{spice}{SPICE}{Simulation with Integrated Circuit Emphasis}

\newacronym{bsim}{BSIM}{Berkeley Short-channel IGFET Model}
\newacronym{cmg}{CMG}{Common Multi Gate}

\glsunset{hd}
\glsunset{sram}
\glsunset{slb}
\glsunset{blb}
\glsunset{fet}
\glsunset{finfet}

\newcommand{\nvm}{\gls{nvm}\xspace}
\newcommand{\fefet}{\gls{fefet}\xspace}
\newcommand{\fefinfet}{\gls{fefinfet}\xspace}
\newcommand{\cam}{\gls{cam}\xspace}
\newcommand{\tcam}{\gls{tcam}\xspace}
\newcommand{\sram}{\gls{sram}\xspace}
\newcommand{\cellsl}{\gls{sl}\xspace}
\newcommand{\slb}{\gls{slb}\xspace}
\newcommand{\ml}{\gls{ml}\xspace}
\newcommand{\hdc}{\gls{hdc}\xspace}
\newcommand{\hd}{\gls{hd}\xspace}
\newcommand{\amem}{\gls{am}\xspace}
\newcommand{\imem}{\gls{im}\xspace}

\newcommand{\mlc}{\gls{mlc}\xspace}

\newcommand{\ngram}{\gls{ngram}\xspace}

\newcommand{\nmos}{nMOS\xspace}
\newcommand{\pmos}{pMOS\xspace}

\newcommand{\ion}{\gls{ion}\xspace}

\newcommand{\vdd}{\gls{vdd}\xspace}

\newcommand{\vth}{\gls{vth}\xspace}
\newcommand{\gnd}{\gls{gnd}\xspace}
\newcommand{\ivg}{$I_{\text{D}}$-$V_{\text{G}}$\xspace}
\newcommand{\ivd}{$I_{\text{D}}$-$V_{\text{D}}$\xspace}

\newcommand{\finfet}{\acrshort{finfet}\xspace}

\newcommand{\csrsa}{\gls{csrsa}\xspace}

\newcommand{\lowvth}{low-\vth}
\newcommand{\highvth}{high-\vth}

\newcommand{\mosfet}{MOSFET\xspace}
\newcommand{\cmos}{CMOS\xspace}

\usepackage[hidelinks]{hyperref}
\usepackage{flushend}

\usepackage{cleveref}
\crefname{enumi}{Step}{Steps}
\crefname{section}{Section}{Sections}
\crefname{subsection}{Section}{Sections}
\crefname{figure}{Figure}{Figures}
\crefname{algocf}{Algorithm}{Algorithms}
\crefname{algorithm}{Algorithm}{Algorithms}
\crefname{algocf}{Algorithm}{Algorithms}
\crefname{table}{Table}{Tables}
\crefname{equation}{Eq.}{Eq.}
\crefname{eqnarray}{Eq.}{Eq.}
\crefname{appendix}{Section}{Sections}

\ExplSyntaxOn
\cs_set_eq:NN \__siunitx_unit_output_number_sep:
  \__siunitx_unit_output_number_sep_aux:
\ExplSyntaxOff
    
\robustify\bfseries
\begin{document}

\title{HW/SW Co-design for Reliable TCAM-\\based In-memory Brain-inspired\\Hyperdimensional Computing}

\author{Simon~Thomann,~\IEEEmembership{Member,~IEEE},  Paul~R.~Genssler,~\IEEEmembership{Member,~IEEE} and 
        Hussam~Amrouch,~\IEEEmembership{Member,~IEEE}% <-this % stops a space
	\IEEEcompsocitemizethanks{%
	\IEEEcompsocthanksitem%
		Simon Thomann, Paul R. Genssler, and Hussam Amrouch are with the Chair of Semiconductor Test and Reliability (STAR), University of Stuttgart, Stuttgart 70569, Germany. E-mail: \{thomansn, genssler, amrouch\}@iti.uni-stuttgart.de. Simon Thomann and Paul R. Genssler have contributed equally. \\
		The developed framework and models are publicly available at \url{https://opensource.mlcad.org} .%
	}%
}

\IEEEtitleabstractindextext{
\begin{abstract}
Brain-inspired hyperdimensional computing (HDC) is continuously gaining remarkable attention.
It is a promising alternative to traditional machine-learning approaches due to its ability to learn from little data, lightweight implementation, and resiliency against errors.
However, HDC is overwhelmingly data-centric similar to traditional machine-learning algorithms.
In-memory computing is rapidly emerging to overcome the von Neumann bottleneck by eliminating data movements between compute and storage units.
In this work, we investigate and model the impact of imprecise in-memory computing hardware, namely TCAM cells, on the inference accuracy of HDC. 
Our modeling is based on \SI{14}{\nano\meter} FinFET technology fully calibrated with Intel measurement data.
We accurately model, for the first time, the voltage-dependent error probability in SRAM-based and FeFET-based in-memory computing.
Thanks to HDC's resiliency against errors, the complexity of the underlying hardware can be reduced, providing large energy savings of up to 6x.
Experimental results for SRAM reveal that variability-induced errors have a probability of up to \SI{39}{\percent}.
Despite such a high error probability, the inference accuracy is only marginally impacted.
This opens doors to explore new tradeoffs.
We also demonstrate that the resiliency against errors is application-dependent.
In addition, we investigate the robustness of HDC against errors when the underlying in-memory hardware is realized using emerging non-volatile FeFET devices instead of mature CMOS-based SRAMs.
We demonstrate that inference accuracy does remain high despite the larger error probability, while large area and power savings can be obtained. 
\textit{All in all, HW/SW co-design is the key for efficient yet reliable in-memory hyperdimensional computing for both conventional CMOS technology and upcoming emerging technologies.}
\end{abstract}

}

\bstctlcite{IEEEexample:BSTcontrol}

\IEEEoverridecommandlockouts

\makeatletter
\def\ps@IEEEtitlepagestyle{
	\def\@oddfoot{\begin{minipage}[t]{1.\textwidth}\footnotesize \centering \copyright 2023 IEEE. Personal use of this material is permitted. Permission from IEEE must be obtained for all other uses, in any current or future media, including reprinting/republishing this material for advertising or promotional purposes, creating new collective works, for resale or redistribution to servers or lists, or reuse of any copyrighted component of this work in other works. DOI: \href{https://doi.org/10.1109/TC.2023.3248286}{10.1109/TC.2023.3248286} \end{minipage}}
	\def\@evenfoot{}
}
\makeatother

\maketitle

\section{Introduction}
\label{sec:intro}

\begin{figure*}[t]
    \centering
    \includegraphics{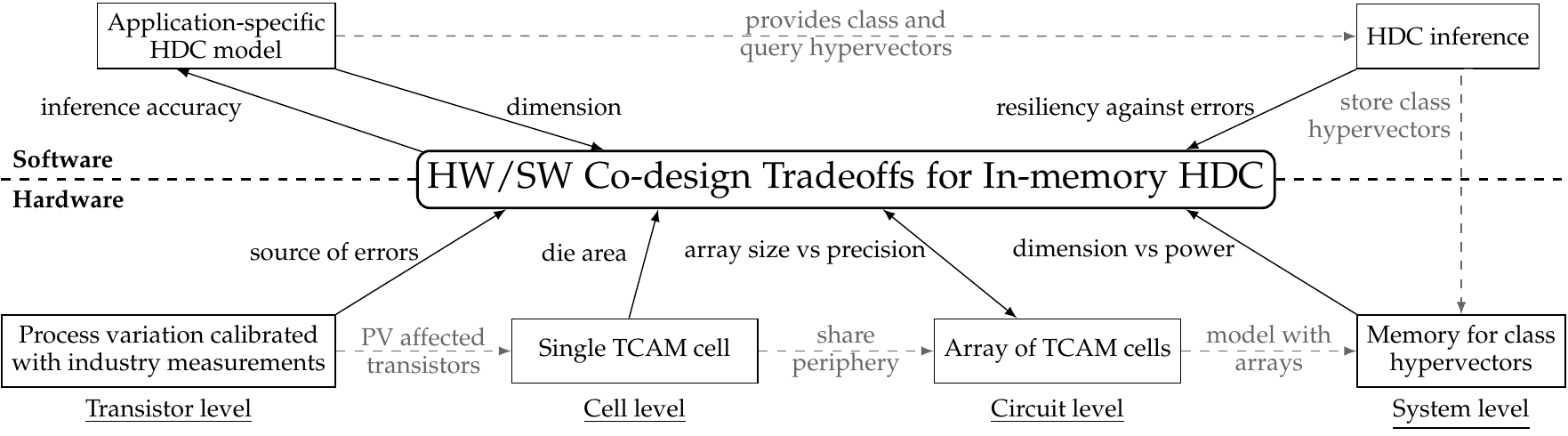}
    \vspace{-0.15cm}
    \caption{Overview of our cross-layer (from transistor to software application) framework for in-memory hyperdimensional computing. It enables the investigation of several tradeoffs at both hardware and software level.}
    \label{fig:overview}
\end{figure*}

\IEEEPARstart{I}{n}
recent years, machine-learning models based on traditional algorithms like \glspl{dnn} have made steady progress.
Such advancements are often associated with larger, more complex neural networks, further increasing their already large demand for processing power and memory.
The traditional von Neumann architecture is reaching its limits for such overwhelming data-centric applications.
As a matter of fact, data movement between memory and compute units profoundly contributes to the total energy consumption~\cite{energyDNN} and does form the key bottleneck.
Furthermore, the significant amount of training data and the iterative training concept to tune the model's weights both exacerbate the energy and processing-power challenges.
Dedicated hardware accelerators, like Google's \glspl{tpu}, aim at accelerating NN inference and training with large power-hungry on-chip memories.
In such accelerators, the data is moved from external memory to the on-chip memories once and then repeatedly feed to the systolic MAC arrays reducing off-chip data movement.
Emerging non-volatile memories consume less energy but are not as reliable as conventional \cmos-based \glspl{sram}.
However, \glspl{dnn} are very susceptible to bit-errors in which a single bit flip can drop the inference accuracy to almost zero \cite{errorsDNN}.

\textbf{Brain-inspired \hdc} addresses the challenges of costly iterative training and the need for large amounts of training data.
Similar to the human brain, patterns can be learned in one iteration from little training data.
For seizure detection, a few samples are sufficient \cite{one2018burrello}.
To create such an \hdc model, only a single pass over the data is sufficient.
This capability is called \textit{one-shot learning} and avoids the costly iterative training required by \glspl{dnn}.
In addition, \hdc operations, unlike in DNNs, do not rely on expensive floating-point matrix multiplications during training but are lightweight and highly-parallelizable bitwise operations.
The same operations are furthermore used for both training and inference, avoiding two distinct hardware implementations.
All of this is achieved through the use of vectors with thousands of independent components -- \textit{hypervectors}.
Hence, an individual component has little impact on the overall accuracy resulting in high robustness against noise in the underlying memory as well as a high resiliency against errors in the performed operations~\cite{rahimi_hyperdimensional_2016, robustness2019manabat, karunaratne_-memory_2020}.
\hdc has been successfully applied to a wide range of application domains such as language recognition~\cite{kanerva_hyperdimensional_2009}, gesture recognition~\cite{rahimi_hyperdimensional_2016}, seizure detection~\cite{one2018burrello}, image classification~\cite{robustness2019manabat}, and more.
In all applications, the inference operation associates an unlabeled query hypervector with the previously-trained class hypervectors by computing a similarity metric between them (e.g., Hamming distance).
The computation is done in the \gls{am}, which stores the class hypervectors.
With traditional computer architectures, the processing unit has to compute the similarity with each class.
\textit{Because of the hypervectors large size (e.g., \SI{10}{\kilo\bits}), data movement is again a bottleneck in existing von Neumann architectures when employed to implement \hdc, especially in regards to applications with a large number of classes.}

\textbf{Novel In-memory Computing} architectures address this bottleneck by implementing the similarity computation directly within the memory where the data resides.
The class hypervectors are not stored in regular memory but in a \tcam~\cite{wuBrainInspired}.
In the traditional address-based memory concept, an address is provided to retrieve data.
In a \tcam, the data is provided, and then its address is returned if the data exists.
This approach can also be used for inference in \hdc.
If the unknown query hypervector is applied as ``data'' to the \tcam, the result is not the address but the similarities to each class.
Therefore, the inference can be fully parallelized, accelerating this step.
In addition, applying the query hypervector is the only data movement reducing the energy costs significantly.
However, the in-memory operations are less precise due to their analog implementation.
The quality of the peripheral circuitry and device-to-device process variation seriously impact the precision, especially in nano-scaled devices (e.g., \SI{14}{\nano\meter} nodes and below).
This holds even more when it comes to emerging memory technologies.
Emerging \glspl{nvm}, like \gls{stt} \cite{basic2013khvalkovskiy}, \gls{rram} \cite{rram2010akinaga} or \fefet \cite{fefet2017duenkel}, promise an increased energy and area efficiency over conventional \sram technology.
Their single-device design require less die area compared to traditional 6T or 8T \sram cells \cite{array1F642019reis}.
Furthermore, emerging \glspl{nvm} consume significantly less power since they are non-volatile and can be turned off without loss of data \cite{array1F642019reis}.
However, these technologies are not yet as mature as CMOS-based \sram and are therefore more affected by variation \cite{sttVariation2020song, rram2021gokul, variation2020ni}.
Nevertheless, \hdc is inherently resilient against errors.
Hence, the software can tolerate imperfect yet more efficient hardware, and HW/SW co-design becomes essential.

\textbf{Our key focus} in this paper is in-memory brain-inspired hyperdimensional computing.
Based on \SI{14}{\nano\meter} \finfet model fully calibrated with Intel measurements for both transistor characteristics and variability, we implement an \sram-based \tcam to compute the similarity metric Hamming distance directly within the memory.
The hypervector is divided into blocks and mapped to individual \tcam arrays.
The computation is less precise in larger blocks and is additionally impacted by process variation.
Thus, we model the probability of error in the Hamming distance computation based on block size and operating voltage in order to explore the available design space and the existing tradeoffs.
Independent of the hardware, at the software level, we model the impact of imprecise distance computations on the inference accuracy for different block sizes.
Finally, we combine both models into a framework to investigate the reliability of in-memory hyperdimensional computing.
Based on this, we explore various HW/SW co-design tradeoffs.

\cref{fig:overview} provides a general overview of our work starting from the transistor level to circuit level all the way up to the software level and the final HW/SW co-design.
With a given set of application-depended class and query hypervectors, our flow enables the exploration of various tradeoffs.
We explore two different applications (language classification and image classification), showing how software matters and HW/SW co-design is key.
As shown, our framework has well-defined interfaces to accommodate different circuit-level models for the similarity metric computation.
Therefore, our work is not limited to a certain  technology node.
Other technologies can be integrated within our framework as long as they offer energy, computation latency, and error probability models.
In this work, we investigate not only conventional \cmos-based \sram, but also emerging \gls{nvm}-based \fefinfet.
The latter promises a higher area and energy efficiency at the cost of a higher probability of error.
To ensure fair comparisons, our \fefinfet device model is based on the same \SI{14}{\nano\meter} \finfet baseline and is subject to the same amount of variations. 

\textbf{Existing works} lack several key aspects.
In \cite{imani_exploring_2017}, a large monolithic array of \SI{45}{\nano\meter} \sram-based XOR gates and counters has been proposed, whereas we utilize advanced \SI{14}{\nano\meter} \finfet \glspl{tcam}.
In combination with the fact that process variation was not considered in \cite{imani_exploring_2017}; hence, no errors could be modeled. 
We employ a calibrated transistor characteristics and variability model that allows us to correctly investigate the impact of process variation on the inference accuracy.
In a proposed \gls{rram}-based crossbar, the process variation model is basic, and its impact is hidden by slower computation~\cite{imani_exploring_2017}.
We demonstrate that in SRAM-based designs, process variation cannot be addressed similarly and instead requires a higher voltage increasing energy consumption.
The tradeoff between energy and inference accuracy in \cite{imani_exploring_2017} is based on a simplified assumption about the impact of errors in \hdc.
We show that the resiliency against errors of \hdc is much larger.
A \fefet-based \tcam design was proposed in \cite{ferroelectric2019ni}.
They consider process variation but only for the older \mosfet design, whereas \finfet is much more susceptible to variation.
In contrast to our work they do not evaluate the impact on the inference accuracy at software level.

\textbf{Distinction from existing work:}
None of the existing works consider detailed cross-layer modeling starting from the transistor level all the way up to the software level in a holistic way.
We demonstrate that HW/SW co-design is the key to efficient yet reliable in-memory \hdc computing.

\noindent\textbf{Our novel contributions within this paper are as follows:}\\
(1) At the software level, we investigate how imprecise similarity metrics computations impact the inference accuracy of different \hdc models. 
We divide the hypervectors into small blocks and limit the largest detectable Hamming distance.\\
(2) At the hardware level, we model the energy consumption as well as the probability of errors in the underlying \sram-based \tcam caused by block size, process variation, and voltage. For accurate modeling, we employ \SI{14}{\nano\meter} \finfet technology fully calibrated with Intel measurement data. Further, we extend our model to additionally account for the \fefinfet technology. This allows, for the first time, a fair comparisons of the reliability of in-memory brain-inspired \hdc when implemented with conventional \glspl{sram} vs. emerging \glspl{nvm}\\
(3) Based on (1) and (2), our HW/SW co-design explores the impact of error-prone \tcam-based similarity computations on the inference accuracy for the first time.
It reveals that HW/SW co-design opens doors to eliminate hardware errors and keep variability effects, which are very challenging in nano-scaled technologies, at bay.
We further explore the tradeoff between energy and inference accuracy under process variation, demonstrating energy savings of 6x through voltage reduction is possible despite the high induced probability of error of up to \SI{39}{\percent}. We demonstrate that \fefinfet has higher losses in the inference accuracy and propose a replica technique to mitigate those losses to the level of \sram.

\section{Hyperdimensional Computing}
\label{sec:hdc}

\hdc has been recently researched in a wide range of application domains~\cite{rahimi_hyperdimensional_2016, one2018burrello, kanerva_hyperdimensional_2009, robustness2019manabat}.
The concept itself was proposed by Kanerva in 2009~\cite{kanerva_hyperdimensional_2009}.
The technique can detect patterns and classify data by mapping real-world entities into \hd space.
Such a space is created by hypervectors with hundreds or thousands of components, including simple bits, integers, real or complex numbers.
While each type of hypervector has its own implementations of four basic operations, the semantics remain the same.
To map complex real-world entities into the \hd space, first, some hypervectors are created randomly to represent basic real-world values.
The individual components of the hypervector are independent.
Hence, an error in one place barely changes the represented real-world value.
In contrast, numbers in traditional binary representation change drastically if a single significant bit is flip.
Thus, \hdc is intrinsically robust against noise and resilient to faults~\cite{rahimi_hyperdimensional_2016, robustness2019manabat, karunaratne_-memory_2020, continual2022karunaratne}.
% The neurons in our brain are not assigned to a specific task only; they instead work cooperatively.
% Similarly, a specific element of a \hd vector does not store a particular piece of information.
% Again, in the \acrlong{hd} space, information is stored using repetition and cooperation.
% This, in turn, makes \hdc intrinsically resilient to noise and errors, which we will explore later on.
Due to the very high dimensionality, the likelihood that two hypervectors are \textit{orthogonal} is very large.
Two binary hypervectors are orthogonal if their normalized Hamming distance is approximately 0.5.
This metric is determined by the number of individual bit pairs that are different between both hypervectors.
In other words, the number of ones in their XOR product.
The resulting number is divided by the length of the hypervectors to normalize it.
For other types of hypervectors, the operation (\textit{measure similarity}) is determined differently but retains its meaning.

Using the generated hypervectors, the complex real-world entity can be encoded into \hd space with the other three basic operations: addition $\oplus$, multiplication $\circledast$, and permutation $\Pi$.
The addition $\oplus$ bundles two or more hypervectors into a single hypervector of the same dimension.
% The output has a similarity of about 0.25 to all of the inputs, hence they are \textit{similar} to each other.
% For binary hypervectors, it is implemented as a bitwise majority operation.
% Each resulting bit is determined by the majority of its corresponding input bits, either there are more zeros or more ones.
The multiplication $\circledast$ binds two or more hypervectors together.
% In binary, this is implemented as a simple bitwise XOR operation.
% The resulting hypervector is orthogonal to all inputs.
The third operation, permutation $\Pi$, operates on a single hypervector and makes it orthogonal to itself. %, in binary, it is a circular bit shift.

\subsection{Language Recognition with HDC}
\label{subsec:language}

To detect the language of an unknown text, not only are the letters' frequencies important but also typical combinations of them, like ``-ing'' or ``-tion'' in English.
The \ngram data structure is able to capture both aspects by encoding $n$ individual letters $L_i$ and their relation to each other through permutation \cite{rahimi_robust_2016}. 
The letter hypervectors $\text{HV}_{A}$ to $\text{HV}_{Z}$ are created randomly once, remain unchanged, and are stored in the \imem.
\begin{equation*}
    \text{HV}_{\text{\ngram}} = \text{HV}_{L0} \circledast \Pi^1(\text{HV}_{L1}) \circledast \ldots \circledast \Pi^{n-1}(\text{HV}_{Ln-1})
\end{equation*}
With a sliding window of size $n$, the whole text is encoded and the \glspl{ngram} are added together to form a single hypervector.
To train the \hd model, for each language (class) a reference text is encoded.
The class hypervector to language mappings are stored in the \amem, the \hd model.
During inference, the unknown text is encoded with the same \ngram technique and the same hypervectors from the \imem.
The encoded hypervector is presented to the \amem as a query. 
The similarities to all stored class hypervectors are computed.
The label (i.e., language) of the class hypervector with the highest similarity is returned as the classification result.

\begin{figure}
    \footnotesize
	\input{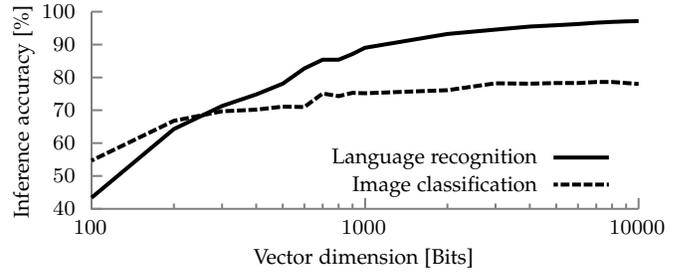}
    \vspace{-0.25cm}
    \caption{The inference accuracy depends on the dimension of the hypervectors and targeted classification problem.}
	\label{fig:dim_ana}
\end{figure}

In this work, we focus on binary hypervectors.
The \ngram encoding is done with $n=4$ as it yields the highest inference accuracy~\cite{rahimi_robust_2016}.
Eight different European languages are each trained with \SI{1}{\million} characters from the Wortschatz Corpora~\cite{quasthoff2006corpus}.
For inference, a query for each language is created with \num{1000} sentences from the Europarl Parallel Corpus~\cite{koehn2005epc}.
The dependency of the inference accuracy on dimension is shown in \cref{fig:dim_ana}.
Increasing the dimension beyond \num{10000} would benefit the accuracy only marginally.
Hence, we limit the investigation to \SIadj{10000}{\bit} hypervectors.

\subsection{Image Classification with HDC}
\label{sec:mnist}

\hdc has been explored for simple image datasets like MNIST \cite{yangVulnerability}, Fashion MNIST \cite{tempGensslerESL}, or character recognition \cite{robustness2019manabat}.
Those datasets have very homogeneous samples without rotations or translations. 
Hence, the \hdc encoding does not need to capture such transformations.
If, for example, images were rotated by \SI{90}{\degree}, the \hdc model has to be retrained to recognize those rotated images.
Another approach was proposed in \cite{searchd}, where multiple class hypervectors are trained and stored to represent a single class (i.e., they have the same label).
For more complex datasets like Omniglot, \cite{karunaratneRobust, mcam} employed a CNN as a feature extractor and \hdc as a classifier.
This approach was extended in \cite{continual2022karunaratne} for CIFAR-100 and miniImageNet.
However, without such preprocessing, image classification with \hdc alone remains a challenge.
 
The MNIST dataset is a common benchmark and does not mandate any processing.
It includes a total of \num{70000} labeled images~\cite{mnisthandwrittendigit2010lecun} for the ten digits from zero to nine.
The images have a uniform size of \num{28}$\times$\num{28} pixels. 
For each of the \num{784} pixel positions, a random hypervector is generated.
The gray-scale images are binarized, and the position hypervectors of all white pixels are bundled into a single image hypervector.
Per class, \num{6000} image hypervectors are selected for training and bundled into a class hypervector.
No further processing, like retraining, is done.
The \amem contains ten classes (digit 0 to 9) and is queried with the remaining \num{10000} image hypervectors.

\subsection{TCAM Implementations for HDC}
Various \tcam implementations for \hdc have been proposed.
They improve on \tcam's inherently low precision argued for in \cite{karunaratneRobust}.
In \cite{mcam}, a 3-bit \fefet multi-bit CAM (MCAM) achieves a similar inference to the PCM-based implementation proposed in \cite{karunaratneRobust}.
Both approaches employ a CNN in the first stage and report results for the Omniglot image dataset.

In \cite{mimhd}, a multi-bit \hdc design (hypervectors with few-bit components instead of 1-bit binary) based on \mlc \fefet is proposed. 
The $k$ class hypervectors are stored in a $k \times 64$ crossbar of MCAM cells originally proposed in \cite{mcam}.
With MCAMs, the distance metric is not the Hamming distance or cosine similarity, but a semi-linear relationship between the distance and the conductance of the MCAM cell.
The \hdc model is retrained to work with the new distance metric.
The impact of process variation is not evaluated. 
Only random bitflips during the encoding and the inference operation are injected, which does not accurately model the variation described in \cite{mcam}.
In our work, we model process variation accurately in the \tcam cells and thus can investigate its impact on the inference accuracy. 

In \cite{searchd}, a pure binary \hdc training and inference was described. 
They use a crossbar of \gls{rram}-based CAM cells to store the class hypervectors.
Each row, which stores a complete hypervector, is connected to a shared wire.
Before an inference operation, this wire is precharged.
When the query hypervector is applied, CAM cells with mismatching bits form a conducting path and discharge the shared wire.
The row with the least mismatches (i.e., the most similar hypervector) discharges the slowest. 
A special CAM sense amplifier detects which of the shared wires discharges the slowest and finishes last.
The hypervector stored in the row with this slowest discharge is the inference result. 
Process variation is considered in their SPICE simulations but only translated to a constant Hamming distance error.
They do not report any incorrect computations because of it.
However, process variation has a higher impact at smaller process nodes, like \SI{14}{\nano\meter} employed in this work compared to \SI{45}{\nano\meter} in \cite{searchd}.

\section{Hardware-level Analysis and Modeling}
\label{sec:hardware_level}

In typical memory, an address is provided to return data.
In \cam, data is provided and, if the data is already present in the memory, its address is returned.
A \tcam is an extension and allows a lookup with ``don't care''.
This lookup capability makes \glspl{tcam} perform best in search operations.
While in typical memory, the data at each address would have to be compared sequentially, in a \tcam the lookup is parallelized.
Hence, latencies within a few clock cycles can be achieved.
\glspl{tcam} are therefore already used in ultra-high performance applications like network switching~\cite{karam2015emerging} or search engine accelerators in databases~\cite{tsai2017search_engine}.
The third ``don't care'' state can be used, for instance, to perform masked IP address lookups~\cite{chang2009a}.
In \hdc, the \amem can be implemented efficiently with a \tcam avoiding costly data movement.
The query hypervector represents the data to search for.
Instead of address lockup, the Hamming distance to all stored class hypervectors can be computed in parallel.
Therefore, instead of \textit{``search''} it is called \textit{``comparison''} in the rest of the paper.

In this section, we first introduce our calibrated CMOS transistor model that includes process variation.
Second, we describe the \sram-based \tcam cell design and, based on it, the block structure to store and compare a part of a hypervector.
Then we use the process variation data from our transistor model to derive the error probabilities for each Hamming distance depending on voltage and block size.

\sisetup{per-mode=symbol}
\begin{figure}
	\centering
	\begin{subfigure}{\columnwidth}
	    \includegraphics[width=\columnwidth]{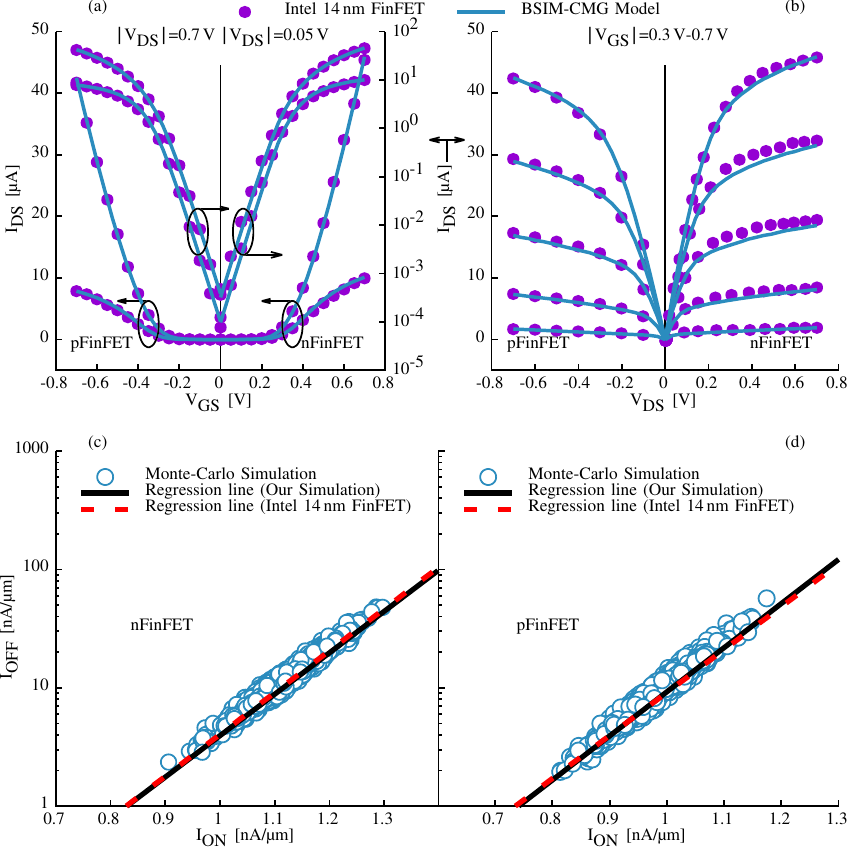}
	    \phantomsubcaption
	    \label{fig:device_char_a}
	    \phantomsubcaption
	    \label{fig:device_char_b}
	    \phantomsubcaption
	    \label{fig:device_char_c}
	    \phantomsubcaption
	    \label{fig:device_char_d}
	\end{subfigure}
	\vspace{-0.3cm}
	\caption{Transistor calibration and validation against Intel \SI{14}{\nano\meter} \finfet measurements extracted from~\cite{intel_data}.
	The industry compact model of \finfet technology (\acrshort{bsim}-\acrshort{cmg}) is calibrated to match Intel data for both \nmos and \pmos transistors for all different biasing conditions (a and b).
	The impact of process variation is also validated against Intel \SI{14}{\nano\meter} \finfet for both \nmos and \pmos transistors in (c) and (d), respectively.}
	\label{fig:decive_charch}
\end{figure}

\subsection{Technology Modeling: \SI{14}{\nano\meter} \finfet Calibration}
\label{sec:technology_modeling}
In this work, we reproduce Intel's \SI{14}{\nano\meter} \finfet measurements~\cite{intel_data} for their mature high-volume manufacturing process.
With SPICE simulations, we carefully tune the transistor model-card parameters for the the industry-standard compact model of \finfet{} (\acrshort{bsim}-\acrshort{cmg}~\cite{7313862}) until they are in excellent agreement with the measurements, both for n\finfet and p\finfet, as demonstrated in  \cref{fig:device_char_a,fig:device_char_b}.
This applies to both, the \ivg and \ivd transistor properties. 
In a second step, we calibrate the model against the measurements for device-to-device variability.
All important sources of manufacturing variability (gate work function, channel length, fin height, fin thickness, and effective oxide thickness) are modeled for a comprehensive representation of the process variation.
Based on the calibrated compact industrial model presented first and Monte-Carlo SPICE simulations, we calibrate the standard deviations for each mentioned source of variability.
\cref{fig:device_char_c,fig:device_char_d} demonstrates that variation in the \gls{ion} vs. \gls{ioff} from our Monte-Carlo SPICE simulations are in excellent agreement with the measurements from Intel.

\subsection{Single TCAM Cell}
\label{sec:tcam_cell}

\begin{figure}[t]
    \footnotesize
    \begin{subfigure}{\columnwidth}
        \phantomsubcaption
        \label{fig:srcam_tcam_cell}
        \phantomsubcaption
        \label{fig:csrsa}
        \phantomsubcaption
        \label{fig:tcam_array}
        \phantomsubcaption
        \label{fig:sa_output}
        \input{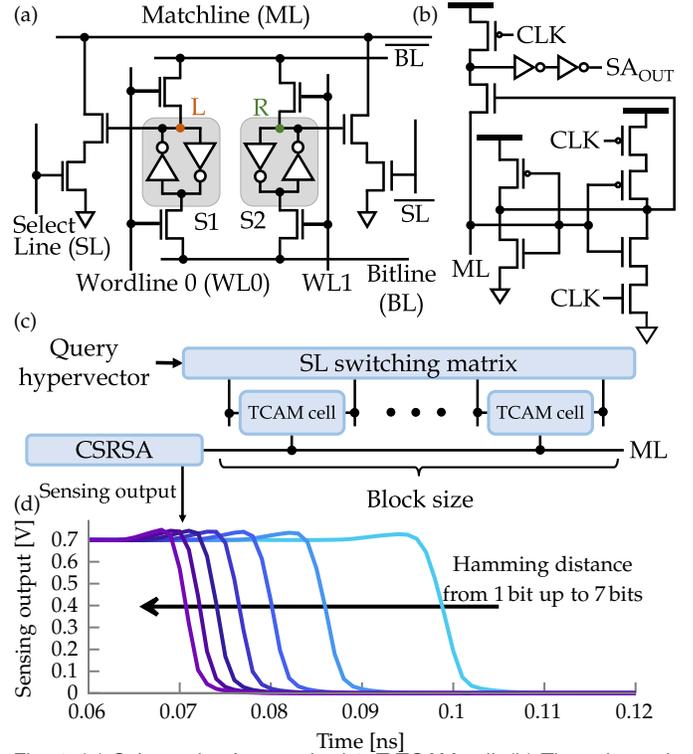}
    \end{subfigure}
    \vspace{-0.25cm}
    \caption{(a) Schematic of a standard 16T \tcam cell.
    (b) The schematic of the \acrlong{csrsa} (\acrshort{csrsa}). The CLK is the enable signal.
    (c) Full \tcam array circuit. The number of cells (block size) is variable.
    Only the select line (SL, input) and match line (ML, output) are drawn.
    (d) Example of the output voltage waveforms of the \acrshort{csrsa} for a block size of \SI{15}{\bits}.}
	\label{fig:tcam_circuit_stuff}
\end{figure}

A single \tcam cell with a 16 CMOS-transistor design couples two SRAMs S1 and S2 as shown in \cref{fig:srcam_tcam_cell}~\cite{ferroelectric2019ni}.
The cell's data $C$, a single bit of a class hypervector, is written to the two SRAMs in a complementary fashion.
For instance, when $C = 1$, S1 and S2 are in the logical $1$ and $0$ state.
A single \sram is a bi-stable element formed by an inverter loop, the labeled nodes (\enquote*{L} and \enquote*{R}) are on the negated side.
The left SRAM S1 holds $1$, but its negated \enquote*{L} node expresses the inverted value $0$.
For a lookup, the \ml is pre-charged.
Then the query data $Q$ is applied to the \cellsl\ (corresponds to the left SRAM S1) and inverted to \slb\ (corresponds to the right SRAM S2).
If it is a \textit{match} ($C = Q$), then the inverted \enquote*{L} and \cellsl are complementary ($\overline{C} \neq Q$) and no conductive path from \gnd to the \ml is formed.
The \tcam cell is OFF because the voltage stays high.
The same logic applies analogously to the right-hand side for the non-inverted \enquote*{R} and inverted \slb\ ($C \neq \overline{Q}$).
If it is a \textit{miss}, either \enquote*{L} and \cellsl or \enquote*{R} and \slb are active at the same time, their associated transistors form a conductive path and discharge the \ml.
The \tcam cell is ON.

\subsection{Our SRAM-based TCAM Array}
\label{sec:tcam_block}

Each individual \tcam cell stores a single bit of a class hypervector.
Multiple cells are combined into a \textit{block} to represent multiple bits of a partial hypervector.
All \tcam cells in the block share the same periphery and access logic as shown in \cref{fig:tcam_array}.
This includes the \gls{bl} and \gls{wl} to write the class hypervector.
In our model, this write operation occurs only during an initialization phase and thus is not within the focus of our framework.
Hence, we exclude it from the evaluation and also simplify the SPICE circuit implementation by replacing the \gls{bl} and \gls{wl} with individual voltage sources.

The \ml is shared with all \tcam cells within one block and an integral part of the \tcam block design.
Before an inference operation, the \ml is pre-charged.
Then the respective bits of the query hypervector are applied to the block.
Recall that each block only represents a few bits of the class hypervector, and thus only the respective bits of the query hypervector are applied.
In each individual \tcam cell, the applied bit and the stored bit are compared as described in \cref{sec:tcam_cell}.
If the result is a miss, the \tcam cell will establish a conductive path from the \ml to \gnd and discharge the \ml.
The more misses occur, the more conductive paths are established, the faster the \ml discharges due to the smaller total resistance.
In other words, the discharge rate reflects the number of cells reporting a miss.
The number of misses between the stored partial class hypervector and the applied partial query hypervector is equal to the partial Hamming distance.
Hence, in practice, a \tcam block realizes the Hamming distance computation.

In our SPICE simulations, the \ml is connected to a \csrsa \cite{ferroelectric2019ni}.
Its schematic is illustrated in \cref{fig:csrsa}.
The \csrsa converts the discharge rate from the voltage domain (how fast is the \ml discharged) to the temporal domain (when does the voltage drop).
This operation latency determines the partial Hamming distance of this block.
An example is shown in \cref{fig:sa_output} for a block size of \num{15} and Hamming distances from one to seven \si{\bits}.
Similar to~\cite{imani_exploring_2017}, we observe a linear dependency between the number of misses and the discharge rate.
While one and two misses are clearly separated by \SI{0.01}{\nano\second}, this gap roughly halves between two and three misses.
The size of the gaps also depends on the size of the block, since large blocks have higher parasitic capacities increasing the discharge time for any number of misses.
Nevertheless, separating large Hamming distances is challenging and impacted by block size and voltage.
Determining the largest detectable distance and attributing a cost in terms of chip area, additional operation latency, or energy is implementation-specific.
Models can be provided for our framework to include such costs into the analysis, extending the explored space.

In this work, we consider \tcam blocks from a size from two \si{\bit} up to \SI{25}{\bits}.
At this upper limit, our evaluation points to a low inference accuracy in conjunction with the maximum precision of \SI{7}{\bits}.
Hence, any larger block sizes would result in low-accuracy systems.
For a full \amem{} (e. g. \SI{10000}{\bits}), multiple block instances are required to store a full hypervector.

\subsubsection{TCAM Block Energy Consumption}
\label{sec:tcam_block_energy}
The block size affects energy consumption.
Larger blocks naturally require more energy but their \tcam cells all share the same periphery; in our circuit, the \csrsa.
Hence, their overhead per bit is smaller.
The energy consumption of a single block is accurately extracted from the SPICE simulations.
All currents over the operation latency of the complete similarity computation are integrated.
The operation latency is the time between the flank of the \csrsa's enable signal and the output voltage dropping below \SI{50}{\percent} \vdd.
It is noteworthy that our framework is modular, and hence any other alternative latency and/or energy models can be included.
In the targeted \SI{14}{\nano\meter} \finfet technology, the nominal voltage is \SI{0.7}{\volt}.
To explore existing tradeoffs between reliability, operation latency, and energy under the effect of voltage, we study a wide range of operating voltages starting from \SI{0.5}{\volt} up to \SI{1.0}{\volt}.

\subsection{Modeling the Error Probability}
\label{sec:erorr_model}

\begin{figure}
    \footnotesize
    \begin{subfigure}{\columnwidth}
        \phantomsubcaption
        \label{fig:variation_histo_15bit_1.0V}
        \phantomsubcaption
        \label{fig:variation_histo_15bit_0.7V}
        \phantomsubcaption
        \label{fig:variation_histo_15bit_0.5V}
	    \input{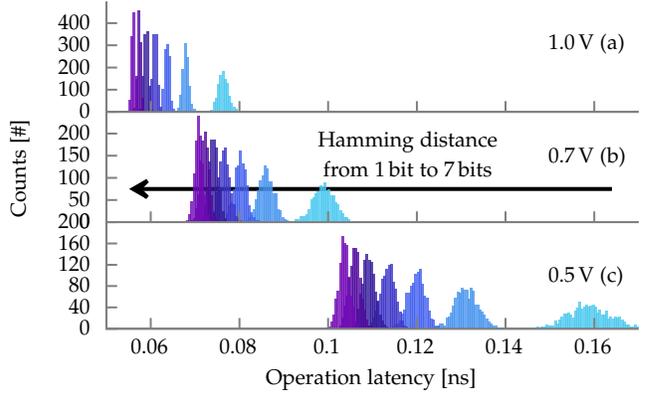}
    \end{subfigure}
    \vspace{-0.5cm}
    \caption{The operation latencies under process variation at three different supply voltages (a, b, and c).
    The block size is \SI{15}{\bits}.
    The results are based on \num{1000} Monte-Carlo SPICE simulations per voltage level and Hamming distance.}
    \label{fig:variation_histo_15bit}
\end{figure}

In this work, the Hamming distance is derived from the operation latency of a block (\tcam cells and \csrsa).
The shorter the latency, the higher the Hamming distance. 
However, the operation latency varies from block to block as shown in \cref{fig:variation_histo_15bit}.
This variation in latency stems from the variation in the underlying transistors. 
Such variation is inherent to every manufacturing process.
It impacts key electrical characteristics of each transistor and hence the reliability and speed of a \tcam cell.
In other words, some cells establish a stronger conductive path from the \ml to \gnd, some cells a weaker path.  
The consequences of these different discharge times of the ML are the different operation latencies. 
Hence, the underlying variation directly impacts the Hamming distance computation and can cause errors.

Our calibrated process variation model (see \cref{fig:device_char_c,fig:device_char_d}), described in \cref{sec:technology_modeling}, is applied to the underlying transistors of a \tcam cell.
We conduct \num{1000} Monte-Carlo SPICE simulations per block size and all possible Hamming distances.
The variation-free nominal operation latencies are shown in \cref{fig:sa_output}, whereas the results of the Monte-Carlo SPICE simulations are presented as histograms in \cref{fig:variation_histo_15bit} for three different voltages.
We analytically model the operation latency distribution for each Hamming distance and build our probabilistic error model.
Without process variation, our framework then maps the nominal operation latency to a Hamming distance (\cref{fig:sa_output}).
With process variation, the framework samples the operation latency distribution (\cref{fig:variation_histo_15bit}) of the nominal Hamming distance.
If the variation from the nominal operation latency is larger than half the distance to the neighboring Hamming distance, then this neighboring Hamming distance is reported as the result of the computation.
Since this is the incorrect distance, it is counted as an error.
As it can be seen from \cref{fig:variation_histo_15bit}, most incorrect Hamming distances are within one bit of the nominal distance. 
Only for large distances (i.e., small latencies), higher deviations are possible. 
The error probabilities for the block size of \SI{15}{\bits} are shown in \cref{fig:error_probability}.
The maximum detectable Hamming distance is defined as seven bits; any shorter operation latency is also reported as seven bits.
While three bits, due to variation, can be incorrectly reported as two or four bits, seven bits cannot be reported as eight bits because of the definition of a maximum detectable Hamming distance.
Thus, the error probability for seven bits is lower.

\subsection{Impact of Operating Voltage}
\label{sec:vdd_impact}

\begin{figure}
    \footnotesize
    \input{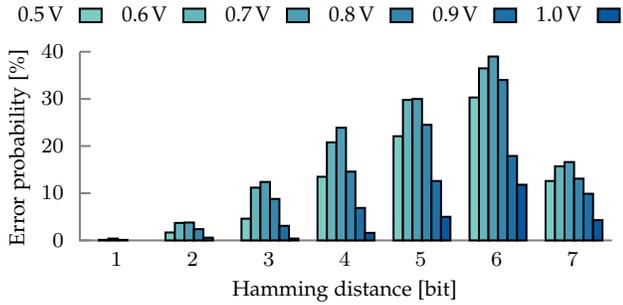}
    \vspace{-0.25cm}
    \caption{Error probabilities of the individual Hamming distances under process variation at different  supply voltages \vdd.
    The error probability is lower for the highest Hamming distance as there is no overlap with an even higher Hamming distances.
    The unintuitive observation of \SI{0.7}{\volt} having the highest error probability is explained in \cref{sec:vdd_impact}.
    The block size is \SI{15}{\bits} and the maximum detectable Hamming distance is seven bits.
    The results are based on \num{1000} Monte-Carlo SPICE simulations per voltage level and Hamming distance.
    }
    \label{fig:error_probability}
\end{figure}

A higher supply voltage \vdd increases the gate voltage at the two transistors on the discharge path.
Thus, the discharge rate is higher and the increase of \ion dominates the linear increase of charge Q of the \ml.
The increase of \vdd also reduces the spread of the individual Hamming distance groups.
As shown in \cref{fig:error_probability}, this directly influences the error probability.
Contrary to the expectations, \SI{0.5}{\volt} performs better than \SI{0.7}{\volt}.
A detailed analysis with finer voltage steps reveals that the interaction between operation latency and spread of the distributions benefits \SI{0.5}{\volt}.
The lower voltage causes wider distributions but those are farther apart.
With \SI{0.7}{\volt}, the distributions are narrower but closer together, resulting in higher overlap and a higher probability of error.
The drawback of the high \SI{1.0}{\volt} level is higher energy consumption.
Even though the discharge rate is higher, the number of charge carriers that have to be moved increases as well.
This increase is directly proportional to the energy and outweighs other reductions.

However, \hdc's resiliency against errors creates the opportunity for energy savings.
If a higher error probability can be tolerated by the \hdc model, than a voltage reduction also reduces the energy consumption.
At \SI{0.5}{\volt}, a \SIadj{15}{\bit} block consumes \SI{0.73}{\femto\joule} per similarity computation in contrast to \SI{4.53}{\femto\joule} at \SI{1.0}{\volt}.
The numbers are derived from our SPICE simulations as discussed in \cref{sec:tcam_block_energy}.
With our HW/SW co-design framework, the tradeoff between inference accuracy and energy consumption is investigated in \cref{sec:codesign}.

\section{Software-level Analysis and Modeling}
\label{sec:software}

At the software level, the inference accuracy for an \hdc model is determined.
Our framework only requires the encoded class and query hypervectors.
Therefore, it is agnostic to the actual application and the encoding of the hypervectors.
Currently, only binary hypervectors are investigated since the underlying hardware models only supports Hamming distance computation.

In-memory Hamming distance computation based on TCAMs/CAMs is restricted to a maximum vector length as shown by our \tcam block design in \cref{fig:tcam_circuit_stuff} and others~\cite{ferroelectric2019ni,imani_exploring_2017}.
Such a block could be, for example, a \tcam array or an \gls{rram} crossbar~\cite{imani_exploring_2017}.
In case of \tcam cell designs, the higher Hamming distances overlap significantly, making their separation difficult as shown in \cref{fig:variation_histo_15bit}. 
Note that the figure does not include the latencies for the Hamming distances from \SIrange{8}{15}{bits} because their overlap makes them indistinguishable in practice. 
In the proposed design, the \gls{tdc} is responsible to differentiate the latencies. 
Its temporal resolution correlates with its implementation complexity.
In other words, a high-resolution \gls{tdc} could differentiate the mean latencies of high Hamming distances.
However, because of the underlying variation, the latencies of high Hamming distances overlap significantly leading to incorrect results in most cases. 
A high temporal resolution has diminishing returns but an increasing implementation complexity.
To avoid this needles complexity of the \gls{tdc}, the detectable Hamming distance is limited by the designer.
This limit is referred to as \textit{precision} in this work.
It is a simplification that aims to reduce the implementation effort of a block by limiting the reported Hamming distance.
As an example, the \SIadj{15}{\bit} block in \cref{fig:sa_output} only reports correctly until seven bits; any higher difference is also reported as seven. 
The \gls{tdc} is emulated by our framework and not part of the SPICE simulations.

Our software-level analysis is orthogonal to the underlying hardware and relies on models.
Initially, a model of error-free and variation-free hardware is assumed.
It accepts a partial class and query hypervector and always computes the correct Hamming distance.
This computation is repeated for all parts of the class and query hypervector and all classes.
The class with the lowest Hamming distance is selected as the inference result.
The result is compared with the true label associated with the test hypervector.
If the inferred class and the true label match, then the inference is correct.
The overall inference accuracy is the ratio of correctly classified test samples to the total number of test samples.
Based on the assumption of an ideal hardware implementation of a block, the inference accuracy is as high as possible for the given \hdc model.

\begin{figure}
    \footnotesize
    \input{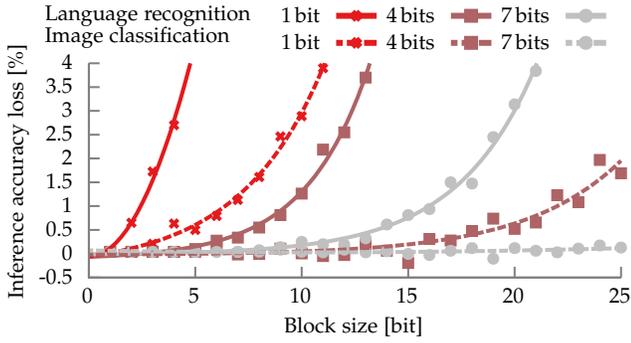}
    \vspace{-0.25cm}
	\caption{Loss in inference accuracy due to different precision levels (in bits).
	Image classification (dashed lines) is more resilient to imprecision than language recognition (solid lines).
	Lines are fitted to data (symbols).}
	\label{fig:accuracy_differences}
\end{figure}

In \cref{fig:accuracy_differences}, different block sizes and precision are shown with the examples of language recognition and image classification.
As a general observation, the inference accuracy starts to drop significantly if the precision of a block is limited to less than half of its size.
This is to be expected since information is lost for larger Hamming distance.
A small distance in another block cannot compensate for that.
Interestingly, in the image classification application, the inference accuracy does not drop as fast.
The reason is the already high overlap of the class hypervector with each other.
On average, their normalized Hamming distance is \num{0.21} in contrast to \num{0.44} for languages as shown in \cref{tab:precision results}.
In other words, the diversity in the image classification model is lower.
Consequently, most queries are similar to the class hypervector and have a lower Hamming distance.
Precision does not limit as much.
This suggests the possibility of a precision-aware encoding scheme to increase the similarities of the classes while at the same time retaining high inference~accuracy.

\begin{figure}
    \footnotesize
    \input{fig8.tex}
	\vspace{-0.25cm}
	\caption{Loss in inference accuracy due to process variation for language recognition using \SI{10000}{\bit}. 
	}
	\label{fig:block_ana}
\end{figure}

Another approach is the generation of quasi-orthogonal hypervectors through metalearning with a CNN as a trainable encoder \cite{karunaratneRobust}.
In \cite{karunaratneRobust},  \num{13180} floating-point hypervector of dimension \num{512} from \num{659} classes were created. 
To align with the previously evaluated classification benchmarks, all hypervectors belonging to a common class are averaged to create a class hypervector.
In addition, all hypervectors are binarized with a simple threshold. 
Their average Hamming distance is 0.49, higher than for the language dataset (0.44) demonstrating the CNNs ability to maximize the distance. 

The baseline inference accuracy with floating-point hypervectors and cosine similarity is \SI{94.3}{\percent}.
After binarization and with Hamming distance, the inference accuracy becomes \SI{93.4}{\percent}.
The results for \sram-based \tcam are summarized in \cref{tab:precision results} and use those binary hypervectors as a baseline.
As expected, without process variation and full precision (block size $N$\,=\,7, precision P\,=\,7), no loss in inference accuracy is reported.
A reduced precision ($N$\,=\,15, P\,=\,7) only impacts the language recognition dataset, Omniglot or MNIST are not affected.
This shows that there is no strong dependency between the average Hamming distance of the class hypervectors and an accuracy loss due to a reduced precision. 
A limited precision, e.g., to half of the block size, limits the detectable Hamming distance to 0.5 (normalized). 
This matches the intended distance to all but the correct class with the quasi-orthogonal prototypes.
The Hamming distance to the correct class is lower and thus less impacted by a reduced precision.

Because of process variation, even a full precision design reduces the inference accuracy by up to \SI{0.28}{\percent}.
The MNIST dataset is again less impacted because a vector size of \num{10000} used in this experiment includes ample redundancy.
In contrast to a design free of process variation, a reduced precision impacts all datasets.

\begin{table*}
    \centering
    \caption{Loss of inference accuracy with and without process variation (PV) and with full precision (block size $N$\,=\,7, precision P\,=\,7) and with reduced precision ($N$\,=\,15, P\,=\,7). The results show that the average Hamming distance between the class hypervectors does not correlate with a requirement for full precision.}
    \label{tab:precision results}
	\sisetup{table-format=1.2}
    \begin{tabular}{lSSSSS}
    \toprule
                &  {Avg. Hamming}  & \multicolumn{2}{|c}{without PV}  & \multicolumn{2}{|c}{with PV} \\
        Dataset &  {Distance} &  \multicolumn{1}{|c}{$N$\,=\,7, P\,=\,7} & {$N$\,=\,15, P\,=\,7} &  \multicolumn{1}{|c}{$N$\,=\,7, P\,=\,7} & {$N$\,=\,15, P\,=\,7} \\
    \midrule
    MNIST       &      0.21   &   0.00\,\%   &   0.00\,\%    &   0.05\,\%    &   0.11\,\%    \\
    Language    &      0.44   &   0.00\,\%   &   0.81\,\%    &   0.28\,\%    &   0.99\,\%    \\
    Omniglot    &      0.49   &   0.00\,\%   &   0.00\,\%    &   0.23\,\%    &   0.31\,\%    \\
    \bottomrule
    \end{tabular}
\end{table*}

\section{HW/SW Co-design and Evaluation}
\label{sec:codesign}

For a holistic analysis of an in-memory \hdc system, models at hardware and software level have to be integrated.
In this section, we demonstrate the promise of HW/SW co-design to obtain an efficient design solution.
The error properties of the underlying hardware are jointly considered with the error resiliency of the target application at the software level.

\subsection{Experimental Setup}
\label{sec:exp_setup}
At the transistor level, we employ our \SI{14}{\nano\meter}\,FinFET model described in \cref{sec:technology_modeling}.
The \tcam cell design and the \tcam block circuit is described in \cref{sec:tcam_cell} and \cref{sec:tcam_block}, respectively.
The error model is developed in \cref{sec:erorr_model}.
We evaluate the inference accuracy with five applications in total.
Language recognition and image classification are introduced in \cref{sec:hdc}.
To confirm our analysis and conclusion, we additionally evaluate EMG gesture detection \cite{latentfactorslimiting2018lobov}, voice recognition ISOLET \cite{ucimachinelearning2017dua}, and heart disease detection CARDIO \cite{ucimachinelearning2017dua}.
The design space includes the voltage levels \SIlist[list-final-separator = {, and }]{0.5; 0.7; 0.8; 1.0}{\volt}, block sizes from two to \SI{25}{\bits}, and an application-dependent subset of dimensions from \SI{2000}{\bits} to \SI{10000}{\bits}.
The hypervectors are partitioned and assigned to the \tcam blocks.
The similarity metrics are computed per block and provided to the error model to capture the impact of process variation.
Additionally, the energy consumption of each \tcam block, which depends on the Hamming distance, is aggregated.
The inference accuracy and energy consumption form one point in the design space.
Any form of preprocessing, such as encoding of data samples, is not considered.

\subsection{Impact of Variability-induced Errors}
\label{sec:results_voltage}

As discussed in \cref{sec:technology_modeling}, process variation is a major concern in advanced technologies such as FinFET.
This is reflected in the probability of errors for a Hamming distance computation is close to \SI{40}{\percent} at the nominal voltage of \SI{0.7}{\volt} (\cref{fig:error_probability}).
However, the inference accuracy decreases at most by \SI{0.45}{\percent} at \SI{0.7}{\volt} (\SI{0.28}{\percent} on average) as shown in \cref{fig:block_ana}.
We have demonstrated in \cref{fig:error_probability} that a reduced voltage of \SI{0.5}{\volt} actually reduces the probability of errors.
This is reflected by an increased inference accuracy.
To further lower the probability of errors, the voltage has been increased to \SI{1.0}{\volt} resulting in a maximum loss of inference accuracy by \SI{0.21}{\percent} (\SI{0.11}{\percent} on average), close to the fluctuations caused by the inherent randomness of any \hdc model.
These results show that \hdc is also resilient against errors in the computation of the similarity metrics.
Although the impact of process variation has to be considered, it does not dominate the tradeoffs available to a system designer.

\begin{figure}
    \footnotesize
	\input{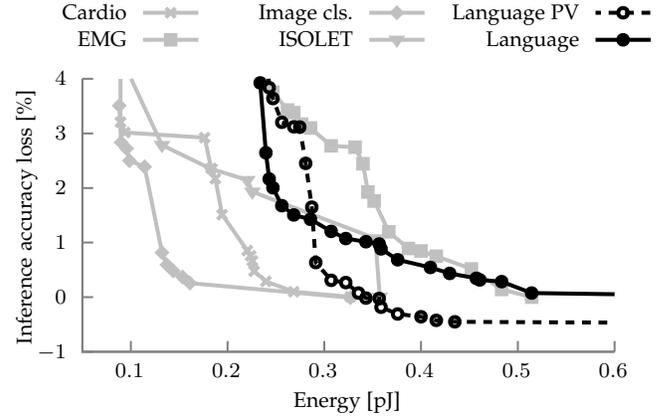}
	\vspace{-0.25cm}
	\caption{Pareto fronts for different hardware configurations.
	A precision level of 7 bits is considered. 
	Impact of Process Variation (PV) is, additionally, demonstrated for ``language recognition''.}
	\label{fig:pareto}
\end{figure}

\subsection{Tradeoff Energy vs. Accuracy}

\cref{fig:pareto} shows the Pareto fronts in the design space for all five applications.
The impact of the application on the HW/SW co-design is discussed in the next section.

\textbf{Voltage} influences the impact of process variation on the inference accuracy as discussed in \cref{sec:results_voltage}.
Our HW/SW co-design shows that the increased energy consumption with \SI{1.0}{\volt} does not outweigh the accuracy gain for almost all configurations.
A higher voltage is only Pareto-optimal if a permille in inference accuracy is required. 
However, due to the random nature of any \hdc model, such small differences are overshadowed by a \enquote{lucky}, or \enquote{unlucky} initialization of the random item hypervectors.
Hence, an increased voltage does not offer a significant benefit and instead offers the possibility to reduce energy consumption.
Accordingly, the large majority of Pareto-optimal configurations use \SI{0.5}{\volt}.

\textbf{Block size} in relation to precision creates an upper bound on the inference accuracy for Pareto-optimal configurations.
For example, to limit the accuracy loss to \SI{2.5}{\percent}, the precision has to be at least half of the block size.
A special case is a precision of one \si{\bit} transforming the \tcam block into a comparator.
Either all bits are equal (Hamming distance is zero) or all are different (Hamming distance is the block size).
To maintain some accuracy, the only viable block size is two.
With \SIadj{10000}{\bit} hypervectors, the loss in inference accuracy is limited to \SI{0.5}{\percent}.
Indeed, for all Pareto-optimal configurations, losses smaller than \SI{0.5}{\percent} require a high level of precision and small block sizes.
In other words, full precision and little to no errors in the similarity metrics computation.
However, process variation is still a source of errors, which have to be tolerated by the \hdc model.

\textbf{The dimension} of the hypervectors creates a similar bound on the inference accuracy. 
The lowest energy consumption is achieved with \SIadj{2000}{\bit} hypervectors.
For language recognition, about a quarter of the Pareto-optimal configurations use this dimension but do not exceed \SI{96}{\percent} accuracy.
A similar portion of the configurations use \SIadj{9000}{\bit} hypervectors and thus achieve the highest accuracy. 
This shows that the inherent resiliency against errors is not purely dependent on a higher dimension as a redundancy buffer.

\subsection{Impact of Application on HW/SW Co-design}

In \cref{sec:software}, the similarities of the class hypervectors to each other are discussed for two applications.
The ten class hypervectors of the image dataset are more similar to each other, whereas the class hypervectors in the language classification model are almost orthogonal.
This difference impacts the resiliency against errors.
The more similar class hypervectors in the image classification make the \hdc model more resilient, as shown in \cref{fig:accuracy_differences}.
The higher resiliency enables more potential for energy savings.
Therefore, block sizes of up to \num{25} are Pareto-optimal for image classification.
Additionally, the impact of process variation is smaller compared to language recognition.
The reason is the smaller Hamming distances of query and class hypervectors.
As shown in \cref{fig:error_probability}, the probability of errors is smaller for lower distances.
Furthermore, the highest Pareto-optimal dimension is \SI{8000}{\bits} for images compared to \SI{10000}{\bits} for languages.
Pareto-optimal configurations for both applications only agree in \SI{0.5}{\volt} and require different parameters otherwise.
These differences highlight the importance of HW/SW co-design.
If a \SI{0.5}{\percent} loss in inference accuracy is acceptable, an efficient implementation of an image classifier saves \SI{4.5}{\siTimes} energy, whereas \SI{11.5}{\siTimes} is possible for languages.

\subsection{Comparison to State of the Art}
In~\cite{imani_exploring_2017}, a \SIadj{4}{\bit} \gls{rram} crossbar is proposed for in-memory Hamming distance computation.
Similar to our work with \sram-based \tcam arrays, the hypervector is partitioned and mapped to the hardware blocks.
In contrast to our approach in which HW/SW co-design is employed to eliminate variability-induced errors, they nullify the impact of process variation by increasing the operation latency.
Therefore, errors only occur if they scale down the voltage of their circuit to reduce energy consumption.
Due to the reduced voltage, the result of all Hamming distance computations is increased by one \si{\bit}.
For example, a distance of two \si{\bits} is always reported as a distance of three \si{\bits}.
If the Hamming distance is four \si{\bits}, five would have to be reported due to the shift.
However, the \SIadj{4}{\bit} \gls{rram} crossbar can only report distances of up to four \si{\bits}.

We implement their error model in our framework.
Thanks to its modularity, we can quickly repeat the analysis analogously and evaluate their error model for different dimensions and applications. In their evaluation, the \gls{rram} crossbar array section that operates at a reduced voltage always causes an error of one \si{\bit} in the Hamming distance.
With the number of scaled voltage \gls{rram} crossbars (i.e., number of bit errors), they assigned an inference loss determined by a theoretical analysis of one application.
In contrast, our framework evaluates the error models by executing the \hdc models.
We show that the \gls{rram} error model does perform better than previously reported.
Their assumption, that each scaled voltage \gls{rram} crossbar causes one bit error, is too pessimistic because the shift applies to all crossbars and thus cancels out.
Instead of \SI{4.0}{\percent}, if all crossbars use a scaled voltage~\cite{imani_exploring_2017}, the accuracy loss amounts to \SI{0.2}{\percent}, which is comparable to the \sram-based \tcam model for a dimension of \SI{10000}{\bits}.
The accuracy loss for other dimensions is comparable as well.
For image classification, the impact of both error models on the inference accuracy is similar.

\section{HDC with Emerging NVM Technologies}
\label{sec:ferro_intro}

\begin{figure}
    \centering
    \begin{subfigure}{\columnwidth}
        \includegraphics[width=\textwidth]{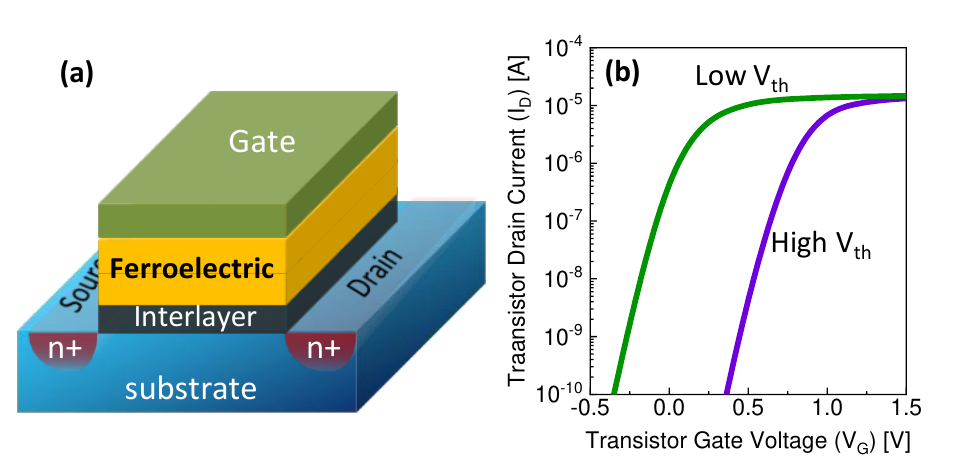}
        \phantomsubcaption
        \label{fig:fefet_device}
        \phantomsubcaption
        \label{fig:fefet_polarization}
    \end{subfigure}
    \vspace{-0.5cm}
    \caption{(a) FeFET where the high-$\kappa$ layer is replaced by a thick (\SIrange{8}{10}{\nano\meter}) layer of ferroelectric material (Hf$_{0.5}$Zr$_{0.5}$O$_{2}$).
    (b) Polarizing the ferroelectric layer in the gate stack creates two distinguishable states, i.e., low \vth and high \vth, which correspond to high and low current, respectively.
    }
    \label{fig:fefet_tcam}
\end{figure}

\begin{figure}
    \centering
    \includegraphics[width=\columnwidth]{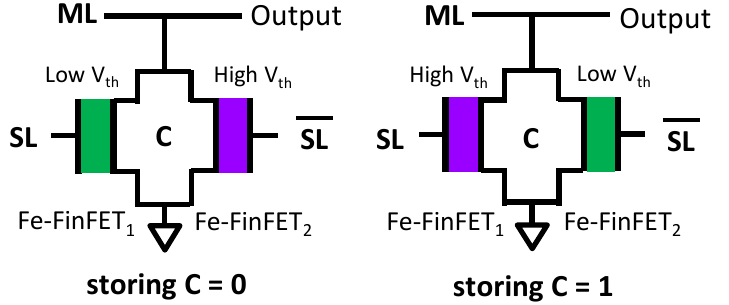}
    \caption{Visualization of the complementary write scheme through \lowvth and \highvth to store logic 0 and 1.
    }
    \label{fig:fefet_write_scheme}
\end{figure}

Data-intensive workloads like big data and deep neural networks demand for larger on-chip memories.
However, traditional \cmos-based \sram consumes lots of power and chip area.
With a slow-down in classical \cmos{} technology scaling, emerging \nvm technologies gather more and more attention.
Several different technologies have been proposed, e.g., \gls{stt}~\cite{basic2013khvalkovskiy}, \gls{rram}~\cite{rram2010akinaga} or \fefet~\cite{fefet2017duenkel}.
\glspl{nvm} reduce the area demand since they do not require six or eight transistors like traditional \glspl{sram} but only the memory device and typically one access transistor~\cite{computing2019reis}.
Furthermore, they retain their state while powered off, reducing static power consumption drastically.
Thanks to these advantages over \glspl{sram}, large on-chip memories are feasible.
However, one major challenge faced by most \glspl{nvm} is their compatibility with the \cmos{} fabrication process, preventing tight integration with logic transistors~\cite{intelSTT2MBl4Cache2019alzate}.

Hafnium Oxide (HfO$_2$) based \fefet addresses this challenge~\cite{fefet2017duenkel,computing2019reis}.
\fefet devices, in fact, are regular \cmos transistors where a thick layer of \gls{fe} material replaces the traditional high-$\kappa$ layer as shown in \cref{fig:fefet_device}.
Existing manufacturing processes can be employed for \fefet~\cite{ferroelectric2018mueller} since HfO$_2$ has been adopted as high-$\kappa$ material since 2007~\cite{hfo2007bohr}.
No additional process steps or new materials are required, only two additional masks~\cite{fefet2017duenkel,fefetIntegration2020beyer}.
Thus, \fefet is the only fully \cmos-compatible emerging \nvm~\cite{fefetIntegration2020beyer,ferroelectric2018mikolajick}.
Additionally, the addition of the FE layer increases only the height of the transistor but not the footprint, enabling a high-density memory.
The FE layer can be polarized with a write operation by applying a \SI{4}{\volt} pulse.
This creates two distinguishable states where the \vth of the \fefet is either high or low as shown in \cref{fig:fefet_polarization}.
The state is sensed by measuring the drain current at a given gate voltage.

Due to all of these advantages, \fefet has a high potential compared to other emerging \glspl{nvm} to be commercialized into mainstream products in the near future. 
Therefore, \fefet attracts interest in academia and industry alike.
GlobalFroundaries has already demonstrated various prototypes~\cite{fefet2017duenkel,fefetIntegration2020beyer} and Intel has recently reported an endurance breakthrough of $10^{12}$ cycles~\cite{fefet_endurance2020Banerjee}.
Furthermore, only two FeFETs are necessary to implement a \tcam cell as shown in \cref{fig:srcam_tcam_cell} making a \fefet-based \tcam array \SI{8}{\siTimes} denser than an \sram implementation~\cite{yin_ultra-dense_2019} creating new tradeoffs.
\subsection{Fe-FinFET-based TCAM Array}
\label{sec:fefet_tcam}

Previous works investigated planar \mosfet-based \glspl{fefet}, which are less susceptible to variation.
In our implementation, we employ the Intel \SI{14}{\nano\meter} \finfet model, also used in the \sram{}-based design, as a base.
To create a \fefinfet, the FE layer is added.
Thus, the underlying \finfet base is complemented by a state-of-the-art ferro transistor physics-based compact model~\cite{ferroelectric2019ni}.
Since we employ the same calibrated model described in \cref{sec:technology_modeling} for the underlying \finfet, we can study and compare the impact of process variation on \fefinfet and \sram in a fair way for the first time.
Our transistor level operation (initializing the polarization, write transistor into desired state) of the \glspl{fefinfet} is similar to recent state-of-the-art shown in~\cite{temperatue_fefet_tcam2021Thomann}.
To program the \glspl{fefinfet}, a \SI{4}{\volt} pulse is applied for \SI{10}{\micro\second} to saturate the polarization for the best read performance.
The state of the \glspl{fefinfet} changes rarely and only if the class hypervectors have to be updated.
Hence, write latency and endurance are not of concern.

In contrast to the 16 \cmos transistor design, a \fefinfet{}-based \tcam{} only requires two \glspl{fefinfet} as depicted in \cref{fig:fefet_write_scheme}.
Both designs use the complementary storing scheme described in \cref{sec:tcam_cell}.
If a logical 0 has been stored in the \tcam, then \fefinfet{}\textsubscript{1} is in the \lowvth{} state (green).
A logical 1 on the \cellsl triggers the discharge and thus signals a miss.
The structure of a full array is analogous to the design shown in \cref{fig:tcam_array}.
Only the \sram{}-based \tcam cells are replace with \fefinfet-based \tcam cells.

\subsection{HDC with Fe-FinFET Under Process Variation}
\label{sec:fefet_results}

\begin{figure}
    \footnotesize
    \begin{subfigure}{\columnwidth}
        \phantomsubcaption
        \label{fig:ferro_variation_histo_15bit_1.0V}
        \phantomsubcaption
        \label{fig:ferro_variation_histo_15bit_0.7V}
        \phantomsubcaption
        \label{fig:ferro_variation_histo_15bit_0.5V}
        \input{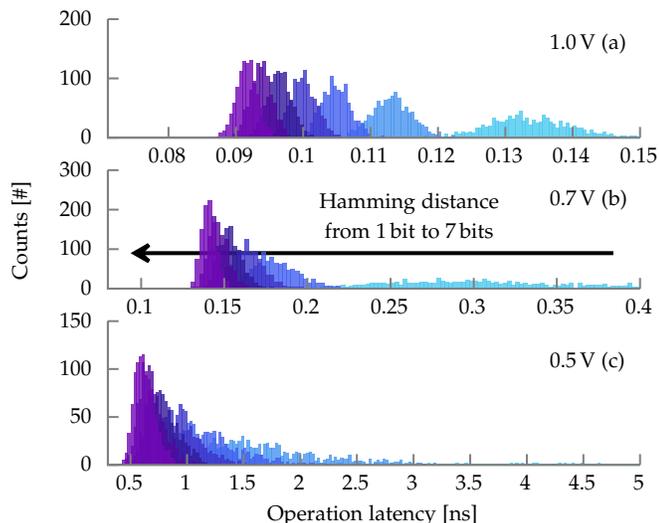}
    \end{subfigure}
    \vspace{-0.5cm}
    \caption{The operation latencies of \fefinfet \tcam under process variation at three different supply voltages.
    The variation for one miss is particularly high at \SI{0.5}{\volt} ranging from \SIrange{0.8}{5.0}{\nano\second} (not recognizable in the plot).
    The block size is \SI{15}{\bits}.
    The results are based on \num{1000} Monte-Carlo SPICE simulations per voltage level and Hamming distance.}
    \label{fig:ferro_variation_histo_15bit}
\end{figure}

\begin{figure}[t]
    \footnotesize
    \input{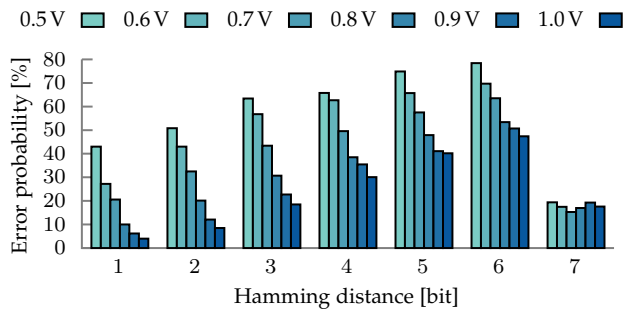}
    \vspace{-0.25cm}
    \caption{\fefinfet \tcam error probabilities of the individual Hamming distances under process variation at different supply voltages \vdd.
    The results are based on \num{1000} Monte-Carlo SPICE simulations per voltage level and Hamming distance.
    The block size is \SI{15}{\bits} and the maximum detectable Hamming distance is seven bits.
    }
    \label{fig:ferro_error_probability}
\end{figure}

\begin{figure}[t]
    \footnotesize
    \input{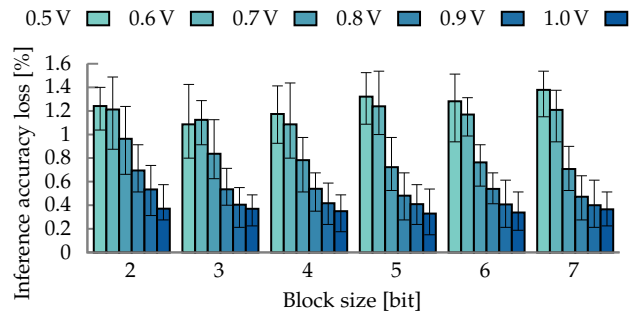}
	\vspace{-0.25cm}
	\caption{Loss in inference accuracy due to process variation on \fefinfet \tcam for language recognition.
	}
	\label{fig:ferro_pv_loss}
\end{figure}

To study an AM with a \fefinfet-based \tcam array, the methodology described in \cref{sec:tcam_block} is applied again.
Comparability is ensured by applying the same capacity to the ML for both designs, \fefinfet and \sram.
The distributions of the operation latencies are shown in \cref{fig:ferro_variation_histo_15bit}.
First, the voltage has a much higher effect on the latency compared to \sram.
\fefinfet is overall slower with up to \SI{5.0}{\nano\second} with one miss at \SI{0.5}{\volt}.
At \SI{1.0}{\volt}, the latency is comparable to \sram at \SI{0.5}{\volt}.
Second, the variation is higher, clearly visible by the large overlap of the latencies representing different Hamming distances.
Similar to \sram, higher voltages do reduce the overlap.
At \SI{0.5}{\volt}, the latencies for a one-bit Hamming distance spread from \SIrange{0.8}{5.0}{\nano\second}.
The sizeable overall overlap is due to \fefinfet{}'s higher susceptibility to process variation compared to \sram.
This is expected since \sram{} is a very mature technology in contrast to \fefinfet{}, which is still in the prototype stage.

\cref{fig:ferro_error_probability} shows that the higher process variation results in a higher error probability.
Whereas \sram peaked at \SI{40}{\percent}, \fefinfet reaches \SI{78}{\percent}.
Interestingly, the observed trend from \cref{fig:error_probability} is not repeated.
Instead, an increase in supply voltage proportionally decreases the error probability.
The worse error probabilities translate further into higher inference accuracy losses.
Although \hdc is very robust against noise, such high error rates inevitably reduce the inference accuracy.
\cref{fig:ferro_pv_loss} shows this increased loss.
For seven-bit block sizes, while in \sram, the loss in inference accuracy can be limited to less than \SI{0.05}{\percent} by using higher voltages, \fefinfet{} can achieve \SI{0.3}{\percent} at best.
For smaller block sizes, inference accuracy loss can be lower but not as low as with \sram.
Energy for a comparison operation is comparable between both technologies with a \SI{19}{\percent} advantage for \sram at \SI{0.5}{\volt} and almost identical at \SI{0.7}{\volt} and \SI{1.0}{\volt} with \SI{+-2}{\percent}.

\begin{table}[b]
      \centering
      \caption{Figures of merit for a single \sram and \fefinfet-based \tcam cell.}
      \label{tab:fefet_sram_numbers}
      \begin{tabular}{lSSS}
            \toprule
              &  {\sram} & {\fefinfet} &  {Diff [\si{\percent}]} \\ 
            \midrule
            Transistors [\#] & 16    & \bfseries 2  & -800  \\
            $E_{mismatch}$ [\si{\femto\joule}] & \bfseries 1.15 &  1.24 & 7 \\
            Latency [\si{\nano\second}] & \bfseries 0.099 & 0.305 &  +308 \\
            \bottomrule
      \end{tabular}
\end{table}

\subsection{Impact of Temperature}
\label{sec:fefet_results temp}
Previous work has investigated the impact of temperature on \fefinfet-based \tcam blocks \cite{temperatue_fefet_tcam2021Thomann}.
They showed that the impact of temperature on \fefinfet-based \tcam is non-intuitive and complex.

To explore its implications at system level, a temperature analysis for both \sram-based and \fefinfet-based \tcam blocks is conducted.
To stay consistent with the previous analyses, the following \tcam configuration is used.
The nominal supply voltage is \SI{0.7}{\volt} and three different temperatures are considered: \SI{27}{\celsius} (baseline), \SI{65}{\celsius}, and \SI{85}{\celsius}.
The temperatures are applied to the whole circuit, i.e., the \csrsa is also affected by the temperature increase.
In addition to the impact of temperature on the underlying FinFET transistor (i.e., reductions in the threshold voltage and carrier mobility), temperature does also degrade the ferroelectric parameters (i.e., remnant polarization $Pr$, coercive field  $Ec$, and saturation polarization $Ps$).

The SPICE analysis captures both effects, the impact of temperature on the underlying FinFET transistor as well as on the ferroelectric layer.
The transistor compact model BSIM-CMG (calibrated with the Intel \SI{14}{\nano\meter} FinFET measurement data) captures the impact of temperature on the underlying FinFET transistor.
The impact of temperature on the ferroelectric layer is additionally modeled by us using experimental data for the P-V loops at various temperatures.
Further details can be found in \cite{guptaTemperatureFerro}.

\begin{figure}
    \footnotesize
    \begin{subfigure}{\columnwidth}
        \phantomsubcaption
        \label{fig:sram_temp}
        \phantomsubcaption
        \label{fig:ferro_temp}
    \end{subfigure}
    \input{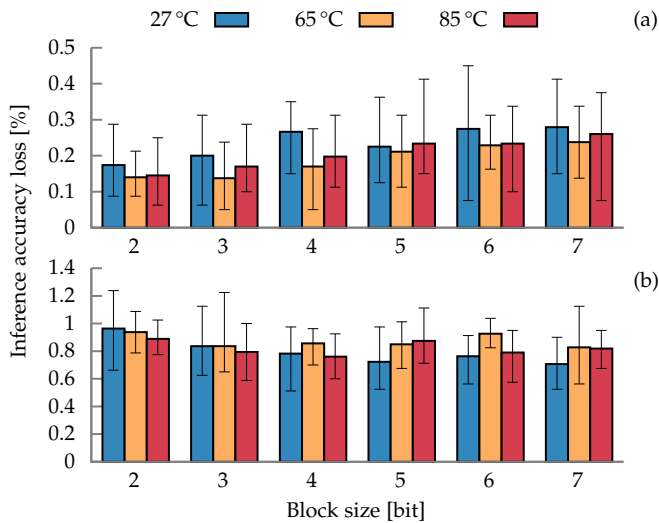}
	\vspace{-0.25cm}
	\caption{Inference accuracy loss on (a) \sram-based and (b) \fefinfet-based \tcam cells due to process variation. For language recognition using \SI{10000}{\bit}, \SI{0.7}{\volt} at different temperatures.
	}
	\label{fig:temp_analysis}
\end{figure}

The results are presented in \cref{fig:sram_temp} for \sram-based \tcam blocks for language recognition at \SI{10000}{\bit}.
For all evaluated block sizes, the inference accuracy loss is lower with higher temperatures.
Two opposing effects are at play. 
On one hand, the temperature-induced increase in operation latency increases the margins between the mismatch groups.
On the other hand, the impact of process variation increases, widening the mismatch groups.
For \sram-based \tcam blocks, the increase in operation latency is dominant and thus the error probability is reduced.
However, this reduction has little impact because the baseline error probability is already high.
In other words, most of the computed Hamming distances are slightly incorrect already and higher temperatures do not correct this significantly.
Only in edge cases, where the error probability is not as high, is the impact of temperature measurable.

For \fefinfet-based \tcam blocks, it depends on the block size which effect has the higher impact as shown in \cref{fig:ferro_temp}.
For smaller block sizes up to \SI{4}{\bit}, a higher temperature reduces the loss in inference accuracy.
With larger block sizes, the increased impact of process variation is higher and thus the loss increases.
The initial investigation suggest possible tradeoffs and challenges at the system level. 
Further experiments and studies are required to explore the potentially positive impact of temperature in the inference accuracy.

\subsection{Perspectives with Fe-FinFET TCAM}
\label{sec:fefet_perspective}

\begin{figure}[t]
    \footnotesize
    \input{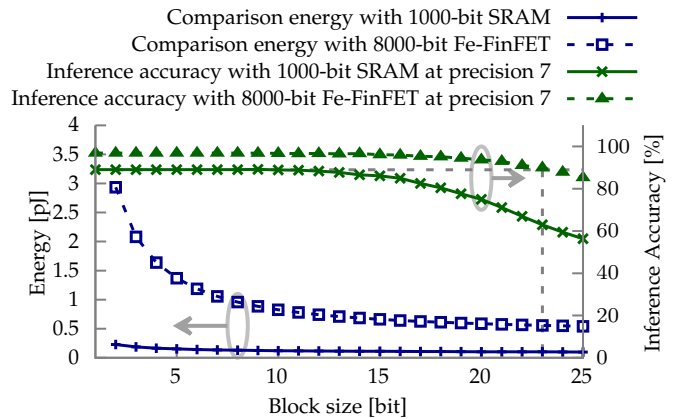}
	\vspace{-0.25cm}
	\caption{With a restricted area budget, more \fefinfet-based \tcam cells are possible resulting in a higher inference accuracy despite the higher variability.
	At a block size of 23, the 8000-bit \fefinfet-based \tcam is as accurate as the \sram baseline and consumes \SI{5.5}{\siTimes} more energy for a comparison of two vectors.}
	\label{fig:area_budget_ana}
\end{figure}

A comparison of the presented results for \sram and \fefinfet underscores the maturity of the \sram process.
Further engineering work at the foundries has to reduce the level of process variation.
Although \fefinfet \tcam{}-based \hdc systems suffer higher accuracy losses, they still offer other benefits.
\Cref{tab:fefet_sram_numbers} highlights the numeric differences.
Most interestingly, only two \glspl{fefinfet} are necessary per \tcam cell.
Even when considering the physical layout, a \fefinfet-based \tcam requires \SI{13}{\percent} of an 16 transistor \tcam{}~\cite{yin_ultra-dense_2019}.
Thus, area-constrained \hdc systems are possible that might otherwise be infeasible with \sram.
Given a limited area budget, an \sram-based system might only fit 1000-bit vectors, whereas a \fefinfet-based system can fit 8000-bit vectors.
This difference in vector length overcompensates the higher process variation.
The \fefinfet-based system achieves a higher inference accuracy, although it consumes more energy for a comparison operation in the AM (see \cref{fig:area_budget_ana}).
Nevertheless, the static power consumption in a \fefinfet-based system is much lower since it can be turned off due to its non-volatility.
To evaluate those savings and compute the total energy consumption, the ratio of comparison operation to idle time has to be known, which highly depends on the target system.

\begin{figure}[t]
    \footnotesize
	\input{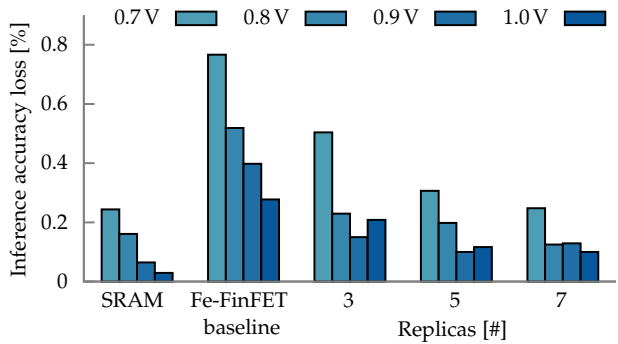}
	\vspace{-0.15cm}
	\caption{Inference accuracy loss comparison of \sram, \fefinfet baseline, and with \fefinfet replicas to reduce the impact of process variation. Results are averaged over block sizes from 2 to 7 bits.
	}
	\label{fig:ferro_redundancy_vs_pv}
\end{figure}

However, such a scheme does not scale if the area budget is sufficient to accommodate a large \sram-based  \gls{am}.
As shown in \cref{fig:dim_ana}, increasing the vector length has diminishing returns.
To still exploit the area savings, the \fefinfet-based arrays can be replicated multiple times.
The query is applied to all arrays simultaneously, thus not increasing latency.
Since all \tcam cells in each array are affected by process variation differently, the arrays can return different Hamming distances.
The median distance is select, effectively ignoring the worst-affected and most inaccurate \tcam cells.
In \cref{fig:ferro_redundancy_vs_pv}, the benefits of this approach are shown. 
The given inference accuracy losses are averaged over all evaluated block sizes.
With seven replicas, the loss in a \fefinfet-based system is on average \SI{0.03}{\percent} higher than in a \sram-based one. 
Such a difference is negligible because the random initialization of the vectors can have a bigger impact.
With three or five replicas, the difference is limited to at most \SI{0.26}{\percent} and \SI{0.09}{\percent}, respectively.
Compared to the \fefinfet baseline, three replicas already reduce the inference accuracy loss by up to \SI{0.29}{\percent}.
Further analyses have to be conducted to evaluate the increased power consumption. 

In summary, for area or power-constrained systems, \fefinfet{} already offers an alternative to \sram.
If the area is not of concern, then \fefinfet-based \tcam arrays can be replicated to reduce the impact of variation.
Advances in manufacturing technology and processes will also reduce this impact, leading to a more mature \fefinfet technology with better performance in \hdc systems, among others.

\section{Conclusion}
\label{sec:conclusion}

In this work, we demonstrated that the resiliency of \hdc against errors is larger than what has been previously assumed.
We revealed the marginal impact of variability on the inference accuracy for \sram-based \tcam cell and the role of applications.
If a \SI{0.5}{\percent} loss in inference accuracy is accepted, up to \SI{11.5}{\siTimes} energy saving would be possible in some applications.
We also investigate the performance of \fefinfet-based \tcam cells and show the effect of process variation on such systems.
Due to its prototype state, inference accuracy losses are higher but chip area and static power can be reduced. 
Such gains can be traded off to counteract the effects of process variation.
\textit{All in all,  HW/SW co-design is a key to achieve efficient, yet reliable in-memory hyperdimensional computing.}

\section*{Acknowledgement}
\addcontentsline{toc}{section}{Acknowledgment}
Authors thank X. Sharon Hu, Xunzhao Yin, and Kai Ni for their valuable support. 
This research was partially supported by Advantest as part of the Graduate School ``Intelligent Methods for Test and Reliability'' (GS-IMTR) at the University of Stuttgart and in part by the German Research Foundation (DFG) ``ACCROSS: Approximate Computing aCROss the System Stack''.

% \vspace{-2em}
\begin{IEEEbiography}[{\includegraphics[width=1in,height=1.25in,clip,keepaspectratio]{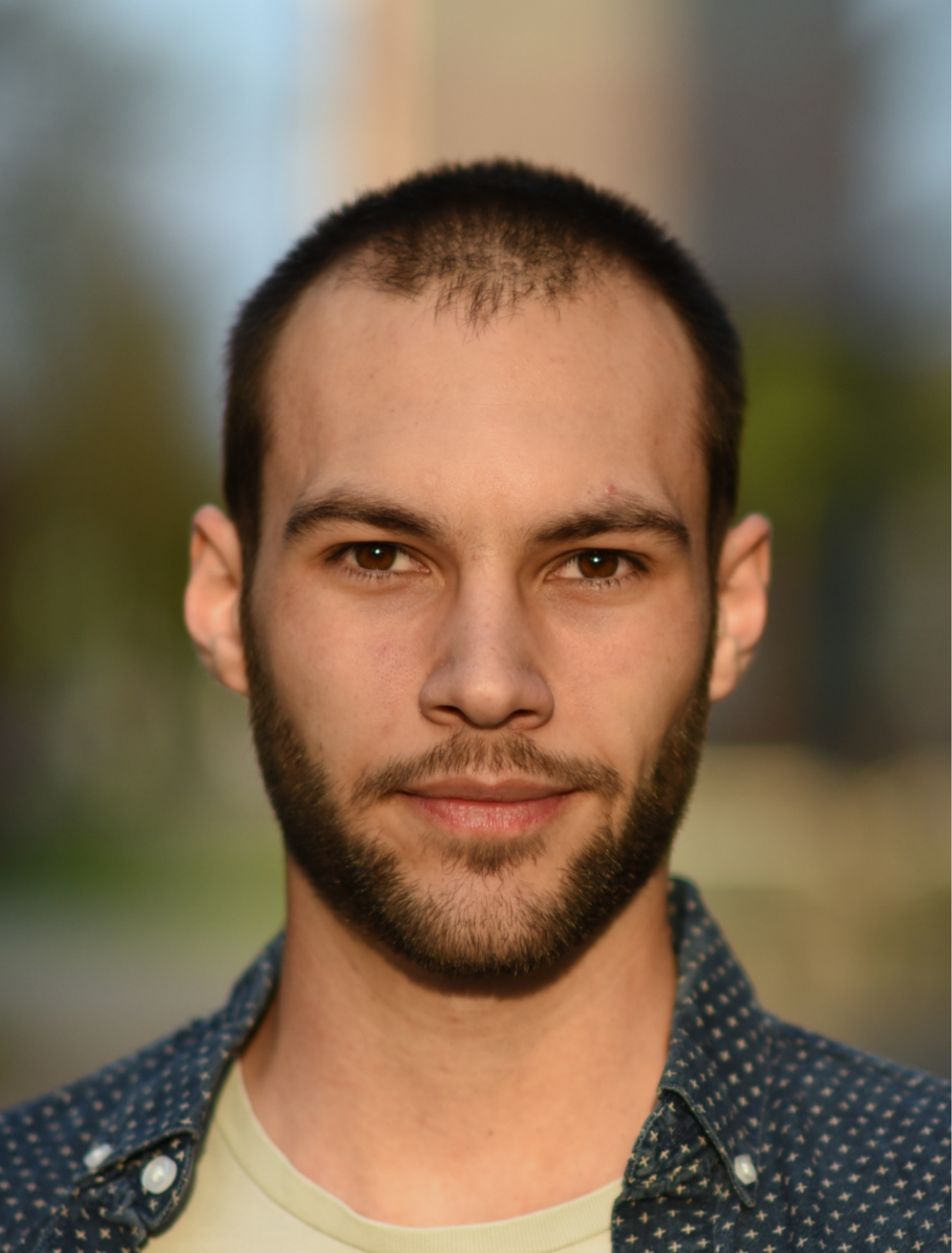}}]%
{Simon Thomann} (M'22) earned his degrees in Computer Science, Master ('22) as well as Bachelor ('19), at Karlsruhe Institute of Technology (KIT), Germany. He is currently pursuing his Ph.D. at the Chair of  Semiconductor Test and Reliability (STAR) at the University of Stuttgart.
His research interests range from device to system level.
His Special interest lies in circuit design, emerging technologies, and cross-layer reliability modeling from device to circuit level.
ORCID 0000-0002-7902-9353
\end{IEEEbiography}

% \vspace{-3em}
\begin{IEEEbiography}[{\includegraphics[width=1in,height=1.25in,clip,keepaspectratio]{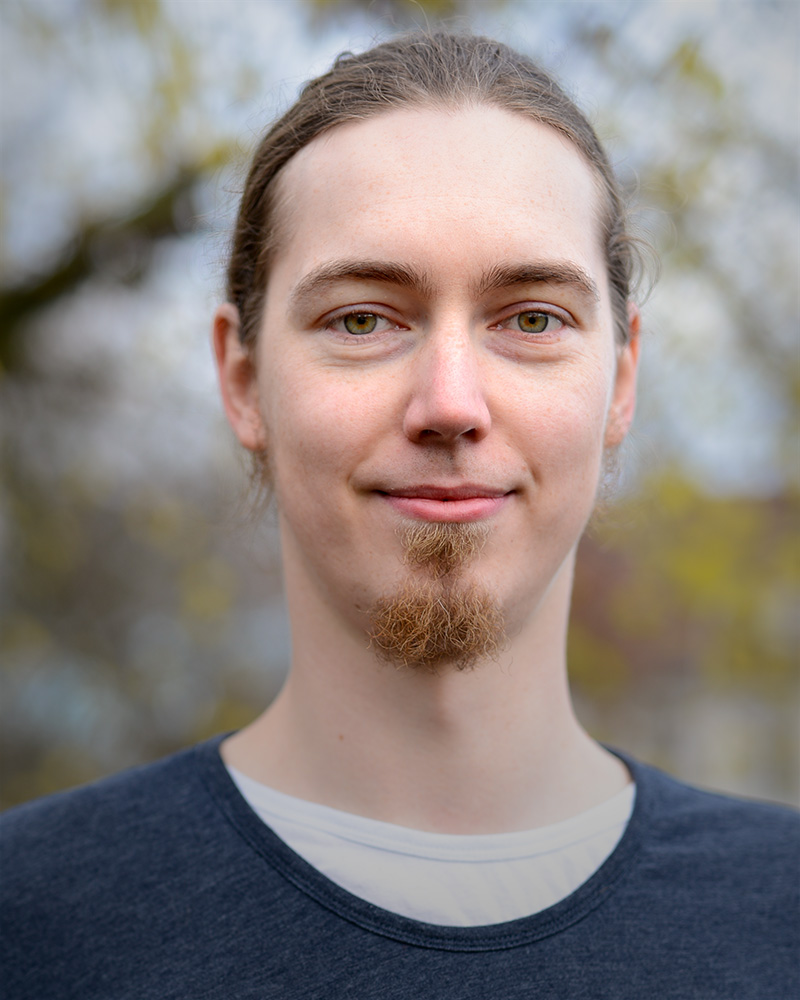}}]{Paul R. Genssler}(M'21) received the Dipl. Inf degree (M.Sc.) in computer science in 2017 at TU Dresden, Germany. In 2018 he started his PhD research at the Chair for Embedded Systems (CES) at Karlsruhe Institute of Technology, Germany. Since 2020 he continues his PhD at the Semiconductor Test and Reliability (STAR) chair within the Computer Science, Electrical Engineering Faculty at the University of Stuttgart. His research interests include emerging technologies, system architecture, and emerging brain-inspired methods for IC test and beyond. ORCID 0000-0002-7175-7284
% \vspace{-3em}
\end{IEEEbiography}
\begin{IEEEbiography}[{\includegraphics[width=1in,height=1.25in,clip,keepaspectratio]{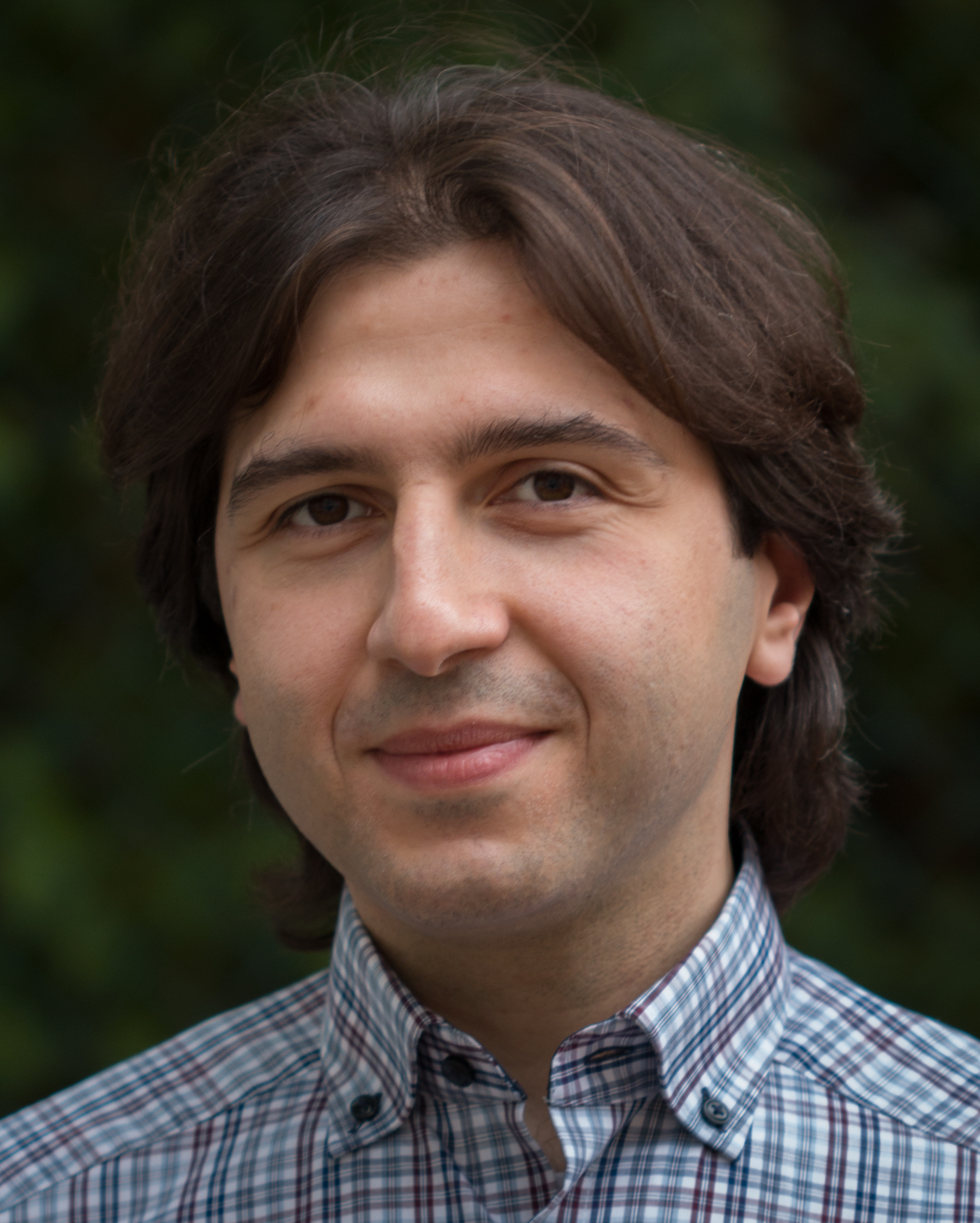}}] {Hussam Amrouch}(S'11-M'15) is a Junior Professor heading the Chair of Semiconductor Test and Reliability (STAR) within the Computer Science, Electrical Engineering Faculty at the University of Stuttgart. He received his Ph.D. degree with distinction (Summa cum laude) from the Karlsruhe Institute of Technology in 2015. 
He holds seven HiPEAC Paper Awards and three best paper nominations at top EDA conferences: DAC'16, DAC'17 and DATE'17 for his work on reliability.
He has 200+ publications in multidisciplinary research areas, starting from semiconductor physics to circuit design all the way up to CAD and computer architecture. Jun.-Prof. Amrouch has delivered 9 tutorial talks in major EDA conferences like DAC and DATE and 25 invited talks (including 2 Keynotes) in several international conferences, universities, and companies. ORCID 0000-0002-5649-3102.
\end{IEEEbiography}

\end{document}